\documentclass[twocolumn,twocolappendix]{aastex6}

\usepackage[utf8]{inputenc}
\usepackage{amsmath}
\usepackage{graphicx}
\usepackage{floatrow}

\newcommand{\dd}[1]{\textrm{d}#1\,}
\newcommand{\Mach}{\mathcal{M}}
\newcommand{\refeq}[1]{Eq.\,(\ref{#1})}
\newcommand{\vct}[1]{\boldsymbol{#1}}
\newcommand{\mtx}[1]{\underline{\underline{#1}}}
\newcommand{\px}{\,\text{px}}
\newcommand{\dg}{^\circ}
\newcommand{\muG}{\mu\text{G}}

\begin{document}

\title{Velocity Anisotropy in Self-Gravitating Molecular Clouds. I: Simulation}
\author{Frank Otto}
\affil{Department of Physics, The Chinese University of Hong Kong}
\email{frank.otto@cuhk.edu.hk}
\author{Weiguang Ji}
\affil{Xinjiang Astronomical Observatory, Urumqi, Xinjiang, China}
\affil{Department of Physics, The Chinese University of Hong Kong}
\and
\author{Hua-bai Li}
\affil{Department of Physics, The Chinese University of Hong Kong}
\email{hbli@phy.cuhk.edu.hk}

\begin{abstract}
The complex interplay between turbulence, magnetic fields, and self-gravity leads to the formation of
molecular clouds out of the diffuse interstellar medium (ISM). One avenue of studying this interplay is by
analyzing statistical features derived from observations, where the interpretation of these features is greatly facilitated
by comparisons with numerical simulations. Here we focus on the statistical anisotropy present in synthetic maps of
velocity centroid data, which we derive from three-dimensional magnetohydrodynamic simulations of a turbulent,
magnetized, self-gravitating patch of ISM. We study how the orientation and magnitude of the velocity anisotropy
correlate with the magnetic field and with the structures generated by gravitational collapse.
Motivated by recent observational constraints, our simulations focus on the supersonic
(sonic Mach number $\Mach \approx 2 - 17$) but sub- to trans-alfv\'enic 
(alfv\'enic Mach number $\Mach_A \approx 0.2 - 1.2$) turbulence regime,
and we consider clouds which are barely to mildly magnetically supercritical
(mass-to-flux ratio equal to once or twice the critical value).
Additionally we explore the impact of the turbulence driving mechanism
(solenoidal or compressive) on the velocity anisotropy. 
While we confirm previous findings that the velocity anisotropy generally
aligns well with the plane-of-sky magnetic field, our inclusion of the effects
of self-gravity reveals that in regions of higher column density, the velocity
anisotropy may be destroyed or even reoriented to align with the
gravitationally formed structures.  We provide evidence that this effect
is \emph{not} necessarily due to the increase of
$\Mach_A$ inside the high-density regions.
\end{abstract}

\section{Introduction}
\label{sec:intro}

It is commonly accepted that the interstellar medium (ISM) is governed by the competition between
the forces of gravity, turbulence, and magnetic fields.  The balance between these forces plays
a major role e.g. for the formation of molecular clouds and of stars (see e.g. \citet{mckee_theory_2007}).
Turbulence is perhaps the most elusive of these forces, because its driving mechanisms are still
not well understood, it is difficult to treat theoretically, and studying it observationally
requires gathering data from a large range of scales. 

Due to the stochastic nature of turbulence, statistical methods are widely used to study
its properties. Many of these methods work with velocity data, because spectroscopic
observations are able to provide such data in plenty, and because active turbulence will
have a direct imprint on the velocity field, in contrast to the density field which may
instead be imprinted with ``fossil'' turbulence.  Notable methods which work
on position-position-velocity (PPV) cubes include the
``spectral correlation function'' \citep{rosolowsky_spectral_1999}, the
``velocity channel analysis'' \citep{lazarian_velocity_2000}, and the
``velocity coordinate spectrum'' \citep{lazarian_turbulence_2004}.
A less elaborate, but still useful method, makes use of the two-point correlation
functions of the velocity centroid \citep{esquivel_velocity_2003, esquivel_velocity_2005, esquivel_velocity_2011, burkhart_measuring_2014}, and we also employ this method in the present work.

While astrophysical turbulence is often assumed to be driven in an isotropic manner,
the presence of magnetic fields introduces anisotropic behavior, because charged particles
are forced to circulate around the magnetic field lines while they may move freely along them.
The size of this effect depends on scale. While large-scale turbulent motions contain more kinetic
energy than small-scale motions, the local magnetic field strength and hence the magnetic energy
density are roughly independent of scale. Therefore the magnetic influence is more pronounced on
small scales, and leads to elongation of the turbulent eddies.
A theoretical model for magnetohydrodynamic (MHD) turbulence in incompressible media was presented
by \citet[hereafter GS95]{goldreich_toward_1995}, and its predictions for the anisotropc scaling behavior
of power spectra have been confirmed by numerical simulations \citep{cho_anisotropy_2000, cho_simulations_2002},
and have been found to also apply for compressible media \citep{cho_compressible_2003}, thus showing
its relevance for studies of the ISM.

Studying this magnetically induced anisotropy from (synthetic) observations of velocity data
has been the topic of a series of studies \citep{esquivel_velocity_2011, burkhart_measuring_2014,
esquivel_studying_2015} based on numerical simulations of MHD turbulence.
It was found that the anisotropy can be reliably detected and used to recover the mean orientation
of the plane-of-sky (POS) component of the magnetic field ($B$-field) for sub- to slightly
super-alfv\'enic conditions ($\Mach_A \lesssim 1.5$), and even if the $B$-field is moderately
inclined ($< 40^\circ$) with respect to the POS.
On the observational side, \citet{heyer_magnetically_2008} applied a directional variant of
principal component analysis to a $^{12}\text{CO}$ PPV cube from the Taurus molecular cloud, and found a significant
correlation between the velocity anisotropy and the mean POS $B$-field orientation.
In a followup study \citep{heyer_trans-alfvenic_2012}, no such anisotropy was found in
the $^{13}\text{CO}$ data which trace the high-density, filamentary region of the cloud.
This difference was attributed to the transition from sub- to super-alfv\'enic turbulence
between the low- and high-density regions.

However, magnetic fields are not the only possible source for velocity anisotropy.
Self-gravity in molecular clouds leads eventually to the formation of large-scale
structures, which are often found to be filamentary, and these oriented structures can have
an imprint on the velocity field of the surrounding medium.  To our knowledge, the effect
of self-gravity on the turbulent velocity anisotropy has not been investigated.

In this study, we present the results from a series of numerical simulations of
driven magnetohydrodynamic turbulence including self-gravity.
Our simulations focus on sub- to trans-alfv\'enic conditions, i.e. where
the magnetic field dominates over or competes with the turbulence. 
This choice is motivated by recent observational constraints
\citep{li_anchoring_2009, li_link_2013, li_link_2014, li_self-similar_2015}
which indicate that the magnetic field is ordered over a large range of scales,
a scenario which is difficult to reconcile with dominating turbulence as this
should lead to rather tangled magnetic field structures.
While many numerical simulations in the past have favored such
super-alfv\'enic scenarios, recent simulations have also started to
explore the sub- to trans-alfv\'enic regime (e.g. \citet{li_magnetized_2015}).

In addition to a range of sonic and alfv\'enic Mach numbers
($\Mach_\text{sonic} = 1.8 - 17.3$ and $\Mach_\text{alfv\'en} = 0.2 - 1.2$),
we explore two different scenarios for the relative strength between
gravity and magnetic forces: one where the simulated volume as a 
whole is in equilibrium between these forces, and
one where the cloud mass is twice the magnetically critical mass.
Again, the choice for these parameters is motivated from recent
observational constraints \citep{li_self-similar_2015}.
Additionally we investigate whether the turbulence driving mechanism,
which may be of more solenoidal or more compressive nature
\citep{federrath_comparing_2010}, has an influence on the velocity
anisotropy.

Our simulation methodology is detailed in Section \ref{sec:simulation}.
From our simulations, we produce synthetic maps of the line-of-sight (LOS) velocity centroid,
and we employ a novel method to detect and quantify the velocity anisotropy from these maps.
(Section \ref{sec:analysis}). We then investigate if and how the anisotropy correlates with the
$B$-field orientation and with the gravitationally formed structures.
(Section \ref{sec:results}). A discussion and comparison
with previous related results is provided in Section \ref{sec:discussion}.


\section{Simulation Method}
\label{sec:simulation}

\begin{table*}[t!]
\centering
\begin{tabular}{rrrrrrrrrrr}
 & & & \multicolumn{4}{c}{solenoidal ($\zeta1$)} & \multicolumn{4}{c}{compressive ($\zeta0$)} \\
ID & $B_0$ & $\dot{e}_\text{drv}$ & $\Mach$ & $\Mach_A$ & AR1 & AR2 & $\Mach$ & $\Mach_A$ & AR1 & AR2 \\
\tableline
 B0E1 &  0 &    10 & 2.01 & $\infty$ &  --  &  -- & 1.90 & $\infty$ &  -- &  --  \\
 B0E3 &  0 &  1000 & 8.67 & $\infty$ &  --  &  -- & 8.65 & $\infty$ &  -- &  --  \\
 B3E1 &  3 &    10 & 1.89 & 0.63     &  4.5 & 8.4 & 1.79 & 0.60     &  5.6 & 7.1 \\
 B3E2 &  3 &   100 & 3.71 & 1.24     &  3.7 & 4.4 & 3.58 & 1.19     &  4.9 & 3.3 \\
B10E1 & 10 &    10 & 2.37 & 0.24     & 10.0 & 5.6 & 2.01 & 0.20     & 10.2 & 2.0 \\
B10E2 & 10 &   100 & 4.06 & 0.41     &  7.8 & 2.7 & 3.73 & 0.37     &  9.8 & 3.9 \\
B10E3 & 10 &  1000 & 8.20 & 0.82     &  3.1 & 4.6 & 7.87 & 0.79     &  3.1 & 1.9 \\
B30E3 & 30 &  1000 & 9.11 & 0.30     &  3.6 & 2.8 & 8.42 & 0.28     &  4.3 & 4.0 \\
B30E4 & 30 & 10000 & 17.3 & 0.57     &  2.8 & 2.8 & N/A  & N/A      &  --  & --  \\
\end{tabular}
\caption{\label{tab:list-of-sims}%
List of simulations and parameters used.
$B_0$ and $\dot{e}_\text{drv}$ are input parameters (see text).
The sonic and alfv\'enic Mach numbers derive from the velocity dispersion
$v_\text{rms}$, which for each simulation is calculated by averaging over a series of
snapshots of fully developed turbulence.  The spread of $v_\text{rms}$
yields an estimate for the relative uncertainty of $\Mach$ and $\Mach_A$ of
about 5\% (at 95\% confidence level).
The AR columns list the maximum aspect ratio of the top-10\% autocorrelation
contour of the column density as encountered during the gravitational stage of
the simulations, AR1 for $cr=1$ and AR2 for $cr=2$; see Section \ref{sec:res-gravity}.
}
\end{table*}

We perform a set of three-dimensional ideal magnetohydrodynamics (MHD) simulations,
using a modified version of the code ZEUS-MP \citep{hayes_simulating_2006}.
The major modification is an implementation of the turbulence driving scheme
described in \citep{stone_dissipation_1998}. While this scheme originally
only used solenoidal (non-compressive) driving modes, we extend it to allow
for an arbitary mix of solenoidal and compressive modes, following
\citep{schmidt_numerical_2006}.  For completeness, mathematical details of
our driving scheme are given in Appendix \ref{app:driving}.

Starting from uniform initial conditions, our simulations proceed in two 
stages. In the first stage, self-gravity is turned off, and turbulence is
driven until the turbulent cascade is saturated (or longer). This is judged
by verifying that the power spectrum of the kinetic energy has reached a
stable distribution.
In the second stage, self-gravity is turned on, while turbulence driving
continues. The gravitational stage is run for several free-fall times, if 
possible, though in the more supercritical scenarios the gravitational
collapse can quickly lead to very high densities, which causes very small
MHD time steps and forces a stop of the simulation.
All the simulations use an isothermal equation of state, and
periodic boundary conditions are employed.

The output of each simulation is a series of snapshot datacubes, each containing
the mass density $\rho(\vct{x},t)$, the velocity field $\vct{v}(\vct{x},t)$, and the
magnetic field vector $\vct{B}(\vct{x},t)$, on a uniform 3D grid of $N^3$ points $\vct{x}$,
at a specific simulation time $t$.
Our simulations are resolved with $N=576$ points per dimension.

The simulations are carried out in reduced units, such that the average mass density
$\rho_0$, the isothermal sound speed $c_s$, and the box length $L$ have unit value.
Turbulence is driven at large wavelengths ($L_\text{drv} = 0.5 L$), and its strength
is controlled by the specific kinetic energy input rate $\dot{e}_\text{drv}$
(measured in units of $c_s^2$).  The magnetic field, if present, is initially uniform,
with strength $B_0$ (in units such that the magnetic pressure is given by $\vct{B}^2/2$)
and pointing in the $z$-direction.  These two parameters, $\dot{e}_\text{drv}$ and
$B_0$, determine the root-mean-square velocity dispersion
$v_\text{rms} = \langle [ \vct{v}(\vct{x}) - \langle \vct{v}(\vct{x}) \rangle_{\vct{x}} ]^2 \rangle_{\vct{x}}^{1/2}$
(where $\langle \rangle_{\vct{x}}$ denotes the spatial average),
and hence set the turbulent sonic Mach number $\Mach= v_\text{rms}/c_s$,
as well as the Alfvén Mach number $\Mach_A = v_\text{rms}/v_A$, where $v_A = B_0/\sqrt{\rho_0}$
is the large-scale Alfvén speed.  
To investigate the influence of the turbulence driving mode, we have carried out
simulations both with purely solenoidal driving ($\zeta=1.0$; see Appendix \ref{app:driving}) and with purely
compressive driving ($\zeta=0.0$).
Table \ref{tab:list-of-sims} lists the simulation
parameters which have been used for the present work.
All our simulations are supersonic, and most of them are sub-alfvénic. 
In the following, we will refer to the simulations by their ID, appended
with ``$\zeta1$'' or ``$\zeta0$'' for indicating whether solenoidal or
compressive driving is used, respectively.

For the gravitational stage, we additionally use the gravitational constant $G$
as a parameter to control the relative strength between gravity and magnetic forces.
The latter stabilize the cloud against collapse if the total cloud mass $M = \rho_0 L^3$ is less than
$M_\Phi = f \Phi \sqrt{4\pi/G}$, where $\Phi = B_0 L^2$ is the magnetic flux through
the cloud, and $f$ is a geometric factor for which we use the value $f=1/2\pi$ \citep{nakano_gravitational_1978}.
Hence the \emph{criticality} parameter $cr = M/M_\Phi$ controls how (magnetically) supercritical the
cloud is, as a whole.  For this work, we have carried out the gravitational stage
of the simulations with both $cr=1$ and $cr=2$.

An example for how our reduced units can be converted to physical units will
be given in Section \ref{sec:discussion}.

\section{Analysis Method}
\label{sec:analysis}

\subsection{Detecting anisotropy strength and orientation}

Our analysis focuses on the velocity centroid, as this is a quantity which is readily
available from observed spectroscopic data.  It is also readily obtained from simulation data
as follows:
for each point $\vct{r}$ on the plane-of-sky, the velocity centroid $V(\vct{r})$ is computed by
integrating the line-of-sight velocity component $v_\text{LOS}$ along the line-of-sight
coordinate $s$, weighted with the local emission intensity, and normalized by the total
emission intensity.  Under optically thin assumptions, the local emission intensity is
proportional to the local gas density $\rho$, so that
\begin{equation}
V(\vct{r}) = \frac{ \int \dd{s} \rho(\vct{r},s) v_\text{LOS}(\vct{r},s) }{ \int \dd{s} \rho(\vct{r},s) }
\quad .
\label{def:vc}
\end{equation}
To identify stochastic anisotropies in this 2D map of velocity centroids,
following \citet{esquivel_velocity_2011},
we look at its two-point second order structure function, which is given by averaging the squared
difference in the velocity centroid between any two points on the map which are
separated by the distance vector $\vct{l}$:
\begin{equation}
SF_V(\vct{l}) = \left\langle \left[ V(\vct{r}) - V(\vct{r}+\vct{l}) \right]^2 \right\rangle_{\vct{r}}
\label{def:sfv}
\end{equation}
This structure function depends on the distance $l$ separating the point pairs,
and on the plane-of-sky (POS) angle $\varphi$ (measured counter-clockwise from the
POS horizontal axis),
\[
SF_V(l,\varphi) = SF_V(l\hat{\vct{e}}_\varphi)
\]
where $\hat{\vct{e}}_\varphi$ is a unit vector with POS angle $\varphi$.

When evaluating \refeq{def:sfv} for a given distance vector $\vct{l}$ in practice,
the average is computed by running over all grid points $\vct{r}$ of the 2D map,
provided that $\vct{r}+\vct{l}$ is also within the map.  Moreover, if $\vct{l}$ is not
aligned with one of the grid axes, the point $\vct{r}+\vct{l}$ will in general not
fall onto a grid point, so that the corresponding velocity centroid value
$V(\vct{r}+\vct{l})$ is not directly available from \refeq{def:vc}; instead, we
compute it by bilinear interpolation from the velocity centroid on the four
neighboring grid points.

To measure the anisotropy present in the structure function $SF_V$, 
we fit (least-squares minimization) its angular behaviour to the following model function:
\begin{equation}
SF_V(l,\varphi) \sim c_l [ 1 - b_l \cos( 2(\varphi - \alpha_l)) ]
\label{eq:fitsfvc}
\end{equation}
This fit is done for each scale $l$, and yields a measure of
the \emph{anisotropy strength} $b_l$ and of the \emph{anisotropy orientation} $\alpha_l$.
We restrict $b_l \geq 0$ and $0 \leq \alpha_l < 180^\circ$ to enforce a unique solution.
The coefficient $c_l = \langle SF_V(l,\varphi)\rangle_\varphi $
measures the average structure function at
scale $l$, and plays no role in determining the anisotropy.

\begin{figure}[tb]
\centering
\epsscale{1.0}
\plotone{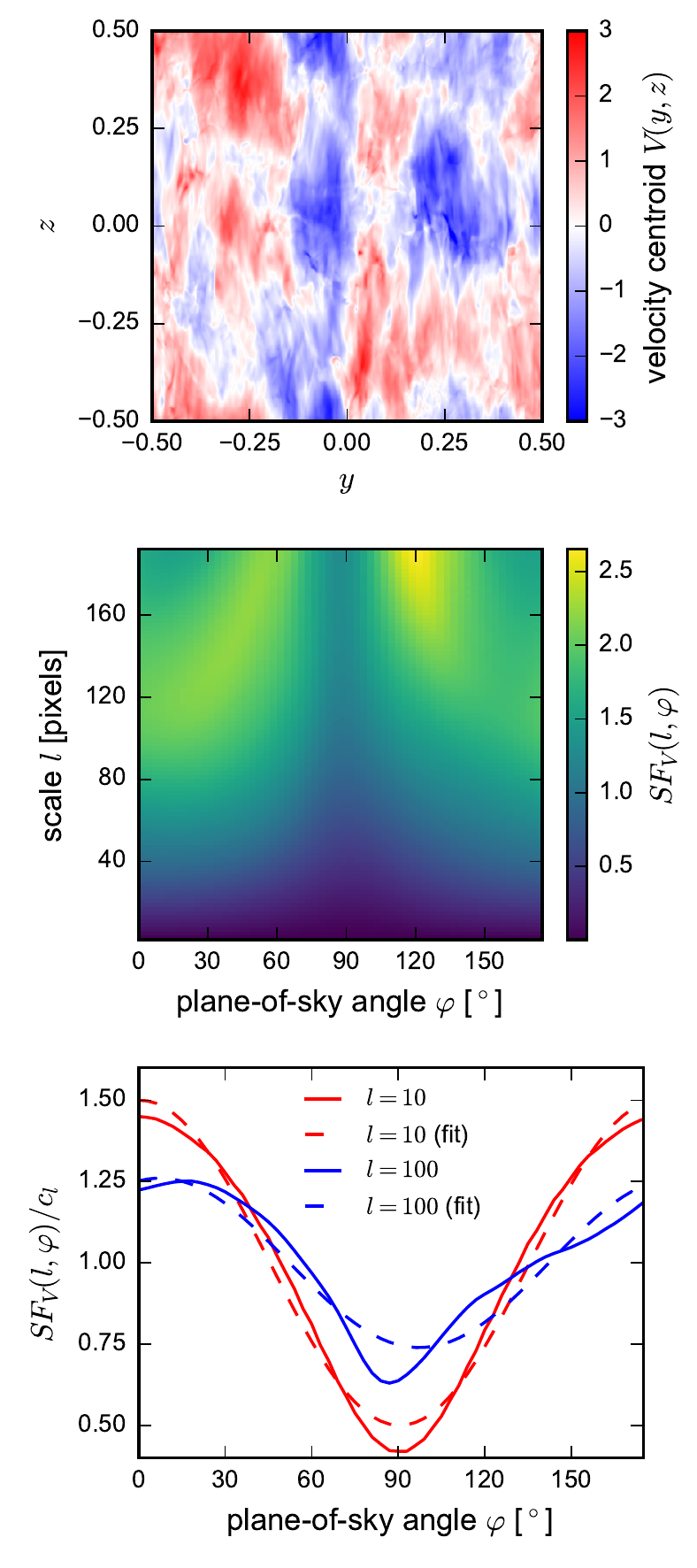}
\caption{(Color online.)
\emph{Top:}
Map of the velocity centroid $V(\vct{r})$ with line of sight along the $x$-direction
(snapshot from the B10E2$\zeta$1 simulation, with saturated turbulence, no gravity).
The mean magnetic field points upwards.
\emph{Middle:}
Map of the structure function $SF_V$ as a function of scale $l$ and
POS angle $\varphi$. At all scales, the structure function exhbits a minimum
around $\varphi=90^\circ$, the direction parallel to the magnetic field.
\emph{Bottom:}
Normalized structure function $SF_V(l,\varphi)/c_l$ plotted against the
plane-of-sky angle $\varphi$. Solid lines show the actual data for the
scales $l=10\px$ (red color) and $l=100\px$ (blue color), while the dotted lines
show the corresponding fits according to the model function \refeq{eq:fitsfvc}.
\label{fig:fitsfvc}}
\end{figure}

Fig. \ref{fig:fitsfvc} illustrates our procedure for obtaining the anisotropy
strength and orientation, using one snapshot of the sub-alfv\'enic simulation
B10E2$\zeta$1.  While the plane-of-sky map of the velocity centroid (top panel)
exhibits anisotropy which can be visually recognized, the picture becomes much
clearer when looking at how the structure function $SF_V$ depends on the scale $l$
and the POS angle $\varphi)$ (middle panel). At all scales, $SF_V$ exhibits a minimum where the
POS angle aligns with the mean POS $B$-field orientation (here, $\varphi \approx 90^\circ$).
When looking at the detailed angular dependence for a single scale (bottom panel),
the approximately sinusoidal angle-dependence of $SF_V$ becomes apparent, and the
position of the fitted sinusoid's minimum informs us about the anisotropy's orientation
($\alpha_l$) while the fitted sinusoid's amplitude yields a measure for the anisotropy's
strength ($b_l$).


\subsection{Parameter uncertainties}
\label{sec:uncertainty}

The fitting procedure described above yields, for each scale $l$, one value
for each of the parameters $\alpha_l$ and $b_l$.  However, the stochastic nature
of the turbulence driving leads to fluctuations of these parameters from snapshot
to snapshot. Even if the structure function is obtained by averaging over a
whole snapshot, there is not enough data in a single snapshot to completely average out these
fluctuations. Therefore it is essential to obtain an error estimate for the fitted
parameters. We propose two methods to do so.

The first method can be applied if snapshots with equivalent physical conditions,
but different realizations of the turbulent field are available.  This is the case
for the first stage of our simulations (i.e. before gravity is switched on):
after the turbulent cascade has saturated,
snapshots which are far enough separated in time are statistically independent,
but correspond to equivalent physical conditions.  Hence we can collect a sample
of the parameters $\alpha_l$ and $b_l$ from individually fitting each snapshot.
This sample is then summarized by the mean anisotropy strength $\bar{b}_l$ and
mean anisotropy orientation $\bar{\alpha}_l$, and the corresponding sample standard
deviations are used as error estimates. (Note that for the orientation, which is
defined on a periodic interval with a period of $180^\circ$, we are using the
\emph{circular mean} and \emph{circular standard deviation}.)
Additionally, the sample size can be doubled by combining equivalent lines of
sight, namely those perpendicular to the mean $B$-field.  We will refer to these
error estimates as \emph{snapshot-sampled} errors.

The second method is designed to be used on a single snapshot. For the second stage
of our simulations, which includes self-gravity, the molecular cloud can collapse,
and physical conditions differ between different stages of the collapse. Hence it
is mandatory to estimate the errors for $\alpha_l$ and $b_l$ from a single snapshot.
Moreover, we are interested in analyzing not only the full POS map, but also selected
subregions of it (e.g. selected by a column density threshold). This is
simply achieved by including only point pairs from the selected region in the
computation of \refeq{def:sfv}.  To obtain an error estimate for these cases,
we randomly cut out from the selected region (or whole map) a set of non-overlapping disc-shaped
areas (of a certain diameter $D$), until no further such area can be cut out;
Each such disc is required to lie contiguously inside the selected region.
For each disc, we then compute the structure function $SF_V$ using only
points inside the disc, and obtain the anisotropy parameters by fitting as above,
yielding values $\alpha_{l,i}$ and $b_{l,i}$ for the $i$-th disc.
These values can be regarded as points $(b_{l,i} \cos(2\alpha_{l,i}), b_{l,i} \sin(2\alpha_{l,i}))$
in a 2D plane, and the geometric center of these points defines the average
anisotropy strength $b_l$ and the average anisotropy orientation $\alpha_l$.
Note that the process of covering the map with non-overlapping discs doesn't cover the
whole map. Therefore it can be repeated several times, and each time a different random
subregion of the map will be covered. For each covering, we obtain different values of
$b_l$ and $\alpha_l$, and again these can be summarized
by the means and the standard deviations. We will refer to the error estimates
obtained in this way as \emph{disc-sampled} errors.

Some technical notes for the disc-sampling error estimates are in order.
First, the shape of disc is chosen so as not to bias the detection of the anisotropy
orientation -- there is no preferred orientation in a disc.
Second, the method is restricted to scales $l$ much smaller than the disc diameter $D$,
because for scales close to $D$, there is an insufficient number of point pairs to
reliably compute a stochastic average of the structure function at that scale -- 
figuratively speaking, there is only a small number of turbulent eddies at larger
scales, which precludes a meaningful statistical analysis.
Third, as the disc-sampling method operates on a single snapshot, it can also be
applied to observational data, where one also doesn't have the luxury of having
multiple snapshots available.

\section{Results}
\label{sec:results}

\subsection{Baseline turbulent anisotropy}

\begin{figure*}[t!]
\plottwo{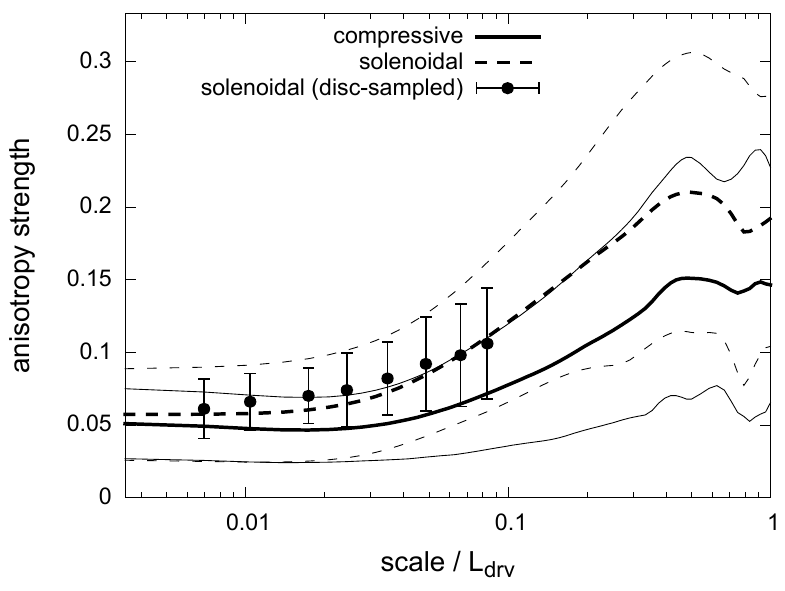}{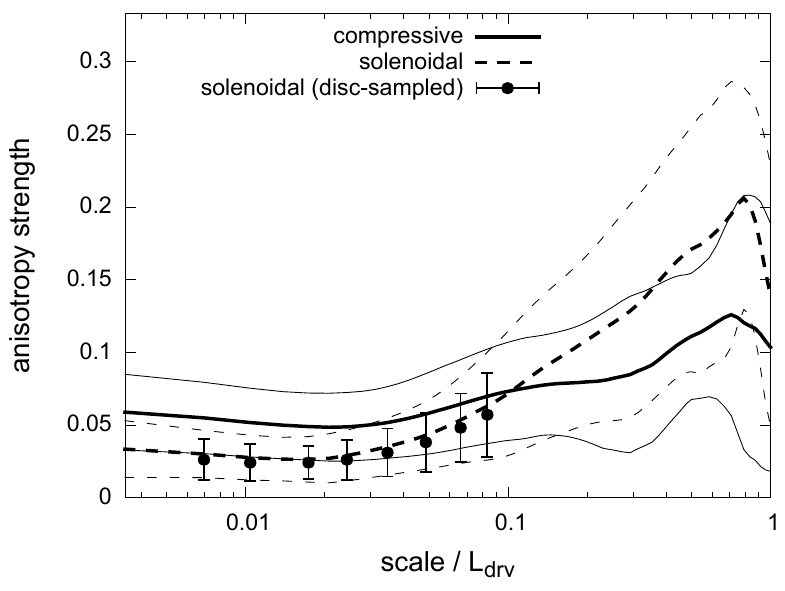}
\caption{%
Scale-dependence of the anisotropy strength $b$ for pure hydro simulations (no $B$-field).
The left panel is for $\Mach \approx 1.95$ (B0E1) while the right panel is for $\Mach \approx 8.66$ (B0E3).
Thick lines show the $b$ data averaged over 3 lines of sight and 10 snapshots of fully developed turbulence.
Thin lines indicate the corresponding $1\sigma$ deviations (snapshot-sampled).
The different turbulence driving mechanisms
(compressive and solenoidal) are represented by the solid and dashed lines, respectively.
The individual data points show the results for solenoidal driving from a \emph{single} snapshot
(with disc-sampled errors; see text).
\label{fig:ai-baseline}}
\end{figure*}

Due to the randomness of the turbulent fluctuations, weak transient anisotropies will
generally appear, even in the absence of magnetic fields or gravity.  To measure the
anisotropy strength of these transient anisotropies, we conducted a set of pure hydrodynamic
simulations (i.e. using $B_0 = 0$) and measured the anisotropy strength of the velocity
centroid map as described in Section \ref{sec:analysis}.
Fig.\,\ref{fig:ai-baseline} displays the snapshot- and LOS-averaged anisotropy strength $b_l$ 
as a function of the scale $l$, for moderate (left panel) as well as highly (right panel)
supersonic turbulence, with both compressive (solid lines) and solenoidal (dashed lines) driving.
The uncertainty of the anisotropy strength has here been estimated by the snapshot-sampled errors,
as described in Sec.\,\ref{sec:uncertainty}.  To verify that the disc-sampled errors give a
similar estimate for the uncertainty, we have evaluated them with discs of diameter $D=64\px$
for a single snapshot, at a few scales up to $l=24\px$.  We find that the disc-sampled errors
tend to be slightly smaller than the snapshot-sampled errors, though both methods agree
reasonably well on the magnitude of the uncertainty.

These results set a \emph{baseline} value for the anisotropy strength.  In the following
investigations, anisotropies will be considered to be non-transient (i.e. caused by other
effects than random turbulent fluctuations) only
if the measured anisotropy strength lies significantly above this baseline strength.

We can make the following observations:
First, the strength of transient anisotropies increases with the scale $l$, reaching a plateau
at the driving scale $L_\text{drv}$. This can be understood by noting that for larger scales,
the number of turbulent eddies at that scale becomes smaller, so that the random orientations
of these few eddies are less likely to average out. On the other hand, at small scales a large
number of eddies is present, and averaging over them leads to a weaker residual anisotropy.
Second, we don't find a conclusive influence of the turbulence driving mode on the
baseline anisotropy strength, as the results agree within their $1\sigma$ deviations.
Third, the results differ very little between simulations with different sonic Mach number.
We find the baseline anisotropy strength to generally lie between $b \approx 0.05$ at small scales and
$b \lesssim 0.25$ at scales near the driving scale. 

\begin{figure}[b!]
\plotone{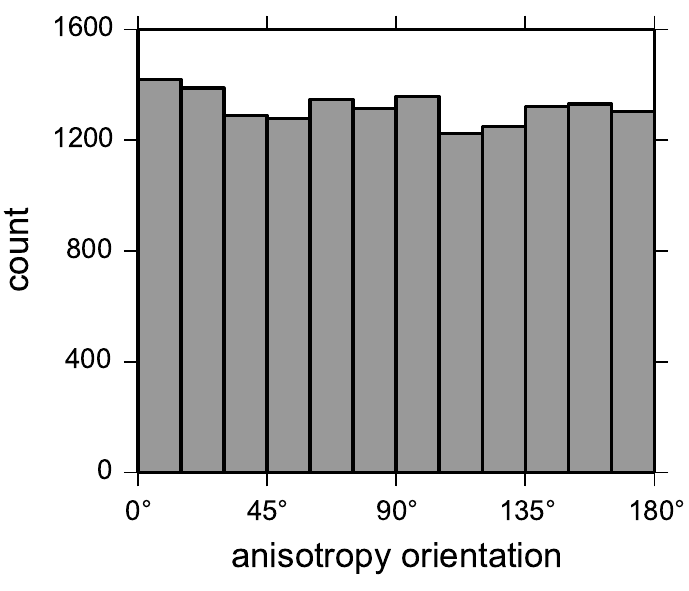}
\caption{%
Distribution of the plane-of-sky angle of the velocity anisotropy at the scale
$l=24\px$, determined for the B0E1$\zeta$1 simulation using disc-sampled ($D=64\px$)
data from 10 snapshots and 3 lines of sight.
\label{fig:hydro-hist}}
\end{figure}

Finally, we note that the orientations of these transient anisotropies are distributed
randomly, as expected in the absence of a governing direction like the magnetic field.
To verify, Fig.\ \ref{fig:hydro-hist} shows the distribution of the measured anisotropy
orientation for one of the pure hydrodynamic simulations.

\begin{figure*}[t!]
\plottwo{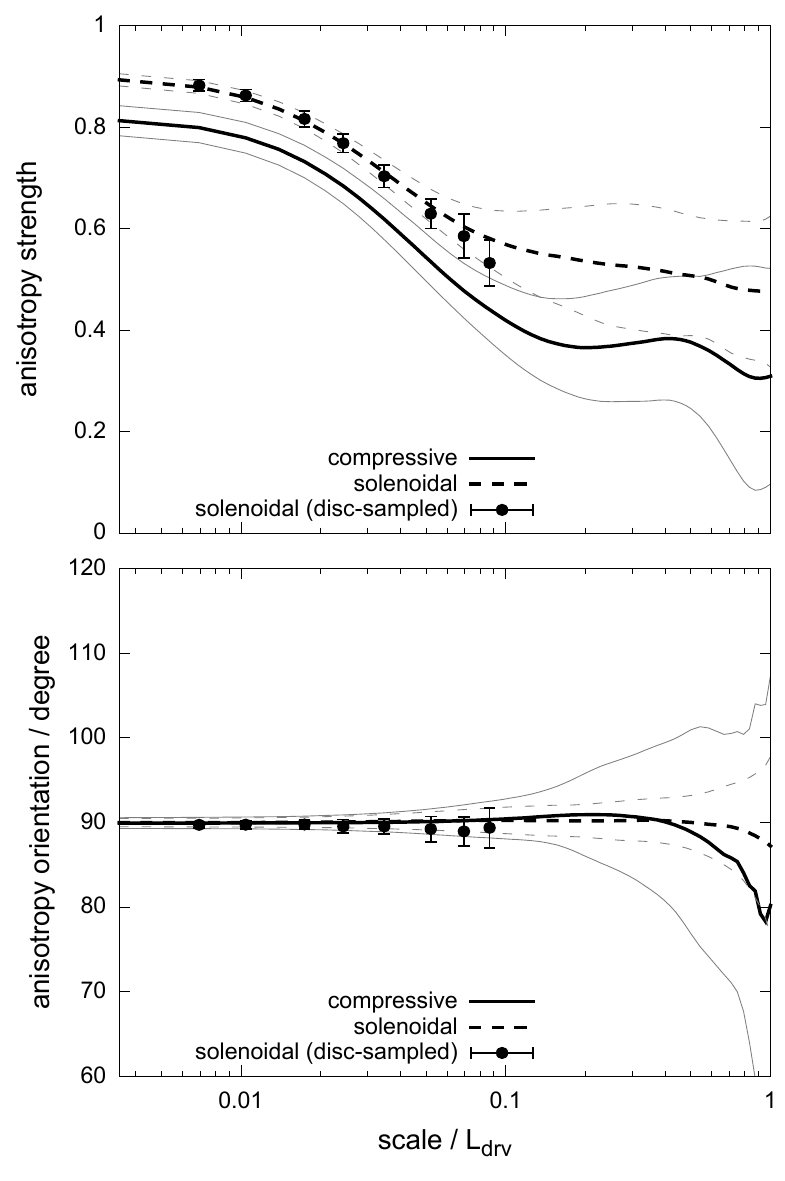}{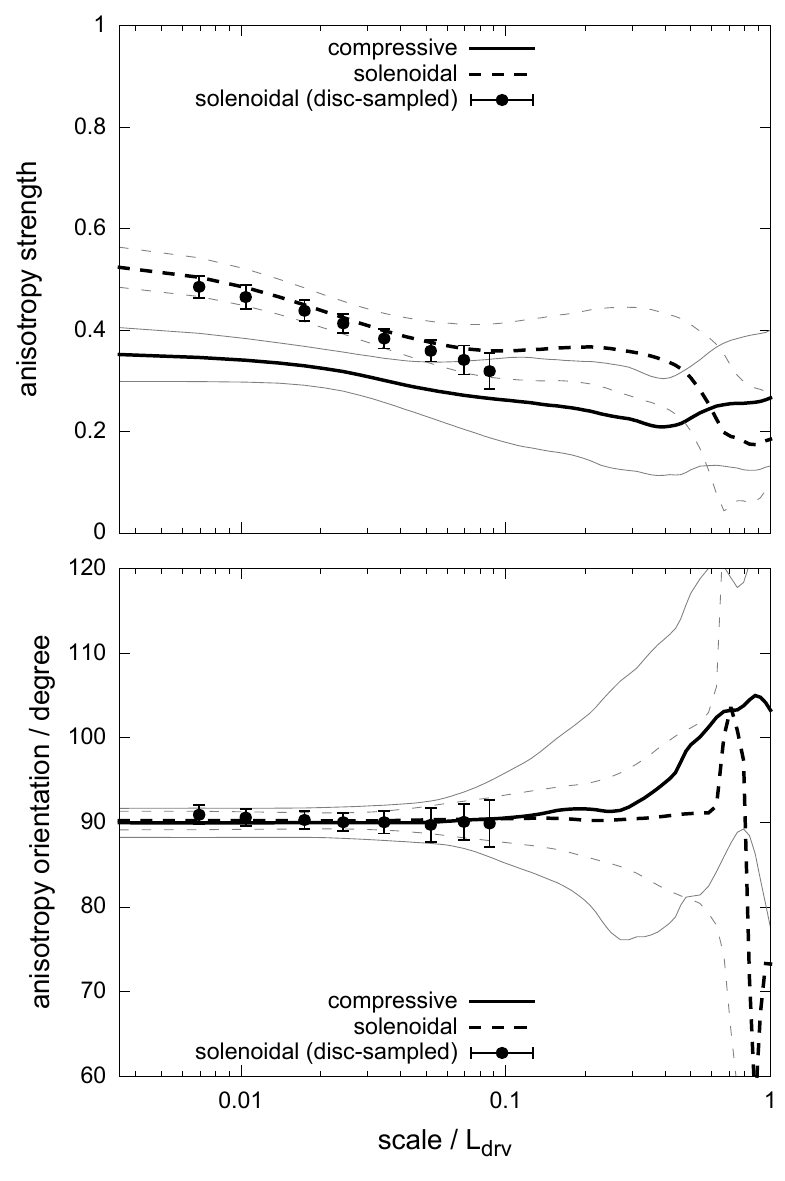}
\caption{%
Scale-dependence of the anisotropy strength (top panels) and the anisotropy orientation
(bottom panels) for strongly sub-alfvénic MHD simulations.
On the lefthand side: $\Mach \approx 2.2, \Mach_A \approx 0.22$ (B10E1);
on the righthand side: $\Mach \approx 8.7, \Mach_A \approx 0.29$ (B30E3).
Thick lines show data averaged over several snapshots of fully developed turbulence,
and averaged over the two lines of sight perpendicular to the mean $B$-field.
Thin lines indicate the corresponding $1\sigma$ deviations (snapshot-sampled).
Solid and dashed lines show data for compressive and solenoidal turbulence driving,
respectively.  As in Fig.\,\ref{fig:ai-baseline}, the individual data points show
the results for solenoidal driving from a \emph{single} snapshot with disc-sampled errors.
\label{fig:ai-subalf}}
\end{figure*}

\subsection{Correlation between anisotropy and $B$-field in absence of gravity}

\placefigure{fig:ai-subalf}

With non-zero $B$-field strength, it is expected from GS95 theory that the
turbulent velocity field will exhibit anisotropy aligned with
the orientation of the mean $B$-field.  Using snapshots from our simulations
where turbulence is saturated, but gravity not yet switched on, we have
measured the anisotropy from the velocity centroid maps where the LOS
is perpendicular to the $B$-field, and indeed we find
that the velocity anisotropy is significantly stronger than the baseline
results from the previous section, and that it aligns very well with the
POS $B$-field orientation. On the other hand, for LOS parallel to the $B$-field,
the velocity anisotropy shows behavior basically indistinguishable from the
baseline ($B=0$) scenario.

The velocity anisotropy is most pronounced for sub-alfv\'enic conditions,
which is exemplified in Figure \ref{fig:ai-subalf} with two simulations
with $\Mach_A < 0.3$.  In the case of mildly supersonic turbulence (lefthand side),
the anisotropy strength is very high ($>0.8$) at small scales and drops continuously
towards the driving scale. Likewise, the anisotropy aligns extremely well (within $1\dg$) with
the POS $B$-field orientation (which is at $90\dg$) at small scales, while at
larger scales the spread in the anisotropy orientation becomes more pronounced,
though even at $1/4$ of the driving scale the spread is still below $10\dg$. 
For strongly supersonic turbulence (righthand side), we see a similar dependence
of the anisotropy strength and orientation on scale, with overall lower levels of
anisotropy strength and larger spreads of orientation. Nevertheless, except
for scales close to the driving scale, the anisotropy strength is significantly
stronger than baseline levels, and the anisotropy aligns very well with the POS
$B$-field, especially at small scales.

\begin{figure*}[tb]
\plottwo{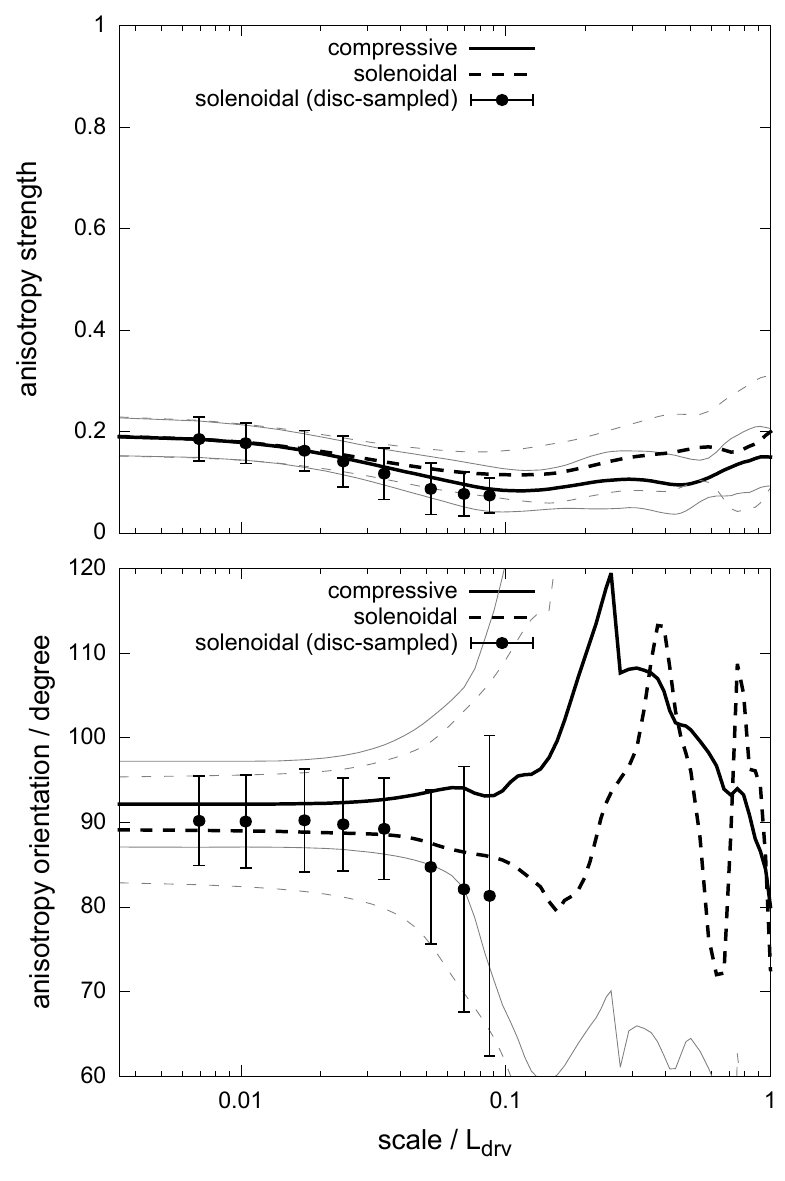}{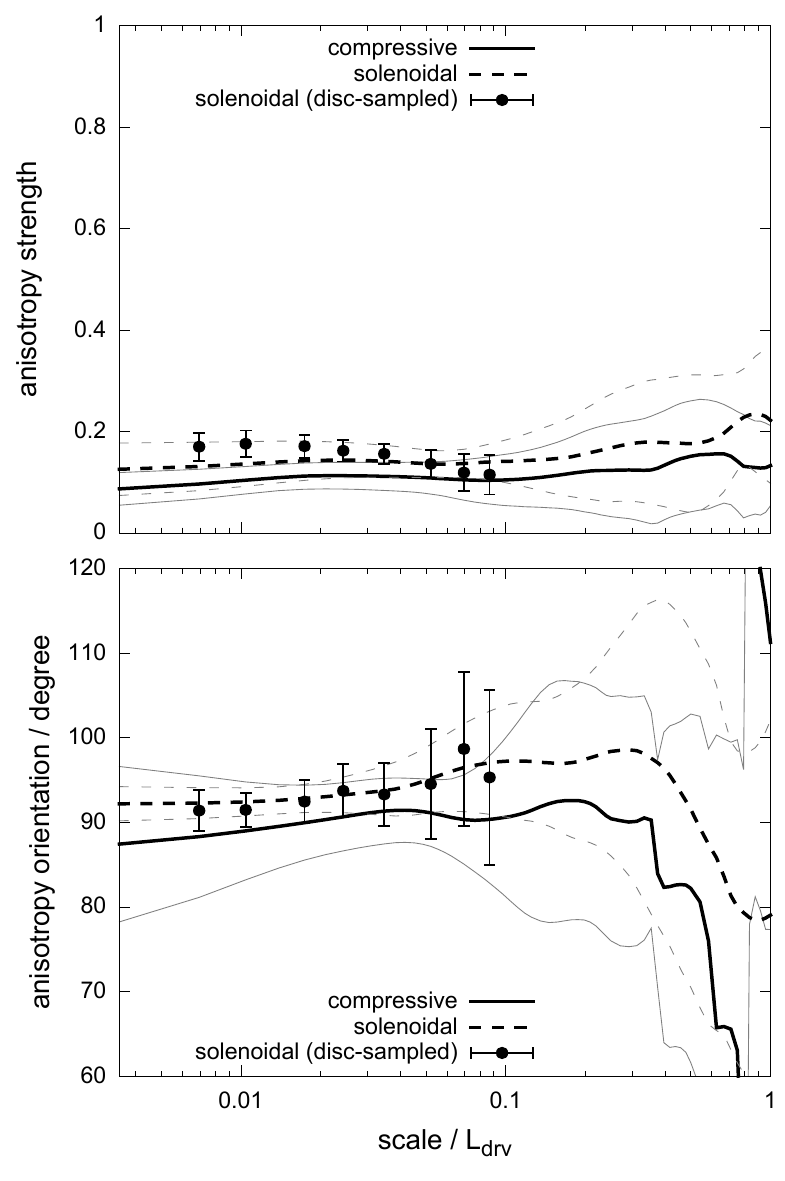}
\caption{%
Like Figure \ref{fig:ai-subalf}, but for trans-alfvénic MHD simulations.
Left: $\Mach \approx 3.6, \Mach_A \approx 1.2$ (B3E2).
Right: $\Mach \approx 8.0, \Mach_A \approx 0.80$ (B10E3).
\label{fig:ai-transalf}}
\end{figure*}

We note that the decreasing anisotropy strength and the increasing misalignment
between anisotropy and $B$-field with increasing scale is in good agreement with
the GS95 theory.  As there is more kinetic energy in large-scale turbulent eddies
than in small-scale ones, at large scales (especially close to the driving scale)
turbulence will be more dominant, and skew the anisotropy results towards the
baseline levels.

Figure \ref{fig:ai-subalf} shows results both for compressive (solid lines) and
solenoidal (dashed lines) turbulence driving.  We observe that solenoidal driving
leads to somewhat stronger anisotropies, which are also somewhat more tightly aligned
with the POS $B$-field up to larger scales. It is possible that the more shock-like
motions, which compressive driving induces, can randomly compress or elongate the
turbulent eddies in random directions, thus having a disorienting effect on the
overall anisotropy.

\placefigure{fig:ai-transalf}

For the case of trans-alfv\'enic conditions ($\Mach_A \approx 1$), the
scale-dependence of the strength and the orientation of the velocity anisotropy
is depicted in Figure \ref{fig:ai-transalf}, with moderately supersonic conditions
on the left and strongly supersonic conditions on the right.  Compared to
the sub-alfv\'enic case, the anisotropy strength is now much smaller
($b \approx 0.1 - 0.2$), but nevertheless it is still elevated
compared to the baseline levels (cf. Figure \ref{fig:ai-baseline}) at
small scales, where also the alignment of the anisotropy with the mean
POS $B$-field is still good (within $\pm 10\dg$).
For larger scales though, the anisotropy strength matches the baseline
levels, and its orientation fluctuates strongly.  Evidently, under these
conditions, the magnetic field is only able to imprint its orientation on
the velocity anisotropy at small scales, where it can dominate over
turbulence.

\begin{figure}
\plotone{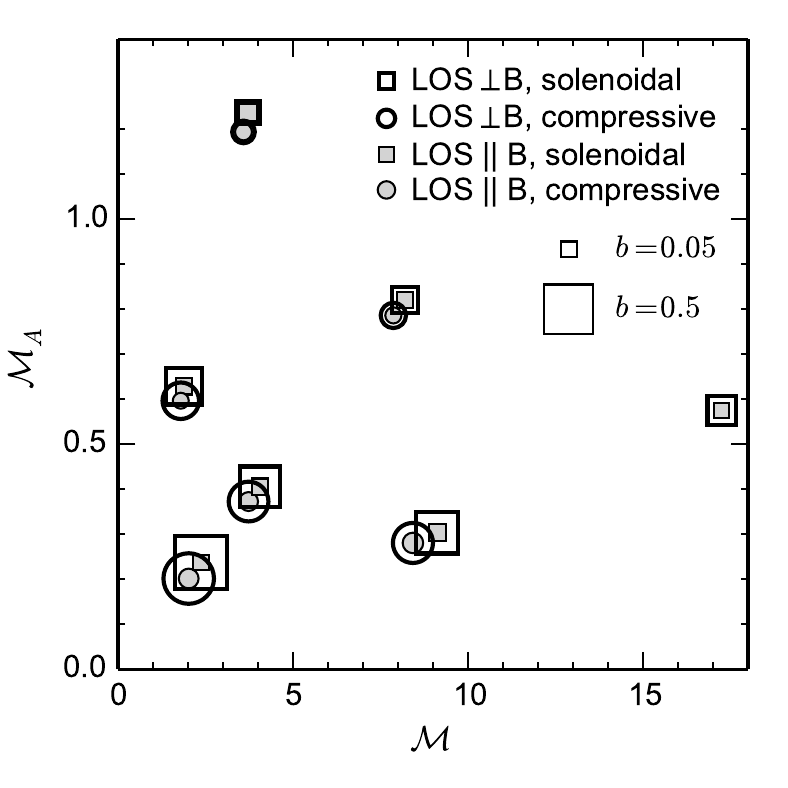}
\caption{%
Anisotropy strength $b$ as a function of sonic ($\Mach$) and alfv\'enic ($\Mach_A$)
Mach number, for saturated turbulence.  The area of each symbol is proportional
to $b$, measured at scale $l=24$ pixels.  Open symbols show
$b$ for LOS perpendicular to the mean $B$-field (averaged over the two possible LOS),
while the grey filled symbols show $b$ for the LOS parallel to the $B$-field.  Squares and circles are used
to indicate solenoidal and compressive driving, respectively.  Each ($\Mach, \Mach_A$)
point corresponds to one of the simulations in Table \ref{tab:list-of-sims}.
\label{fig:ai-all}}
\end{figure}

\begin{figure*}[t]
\plottwo{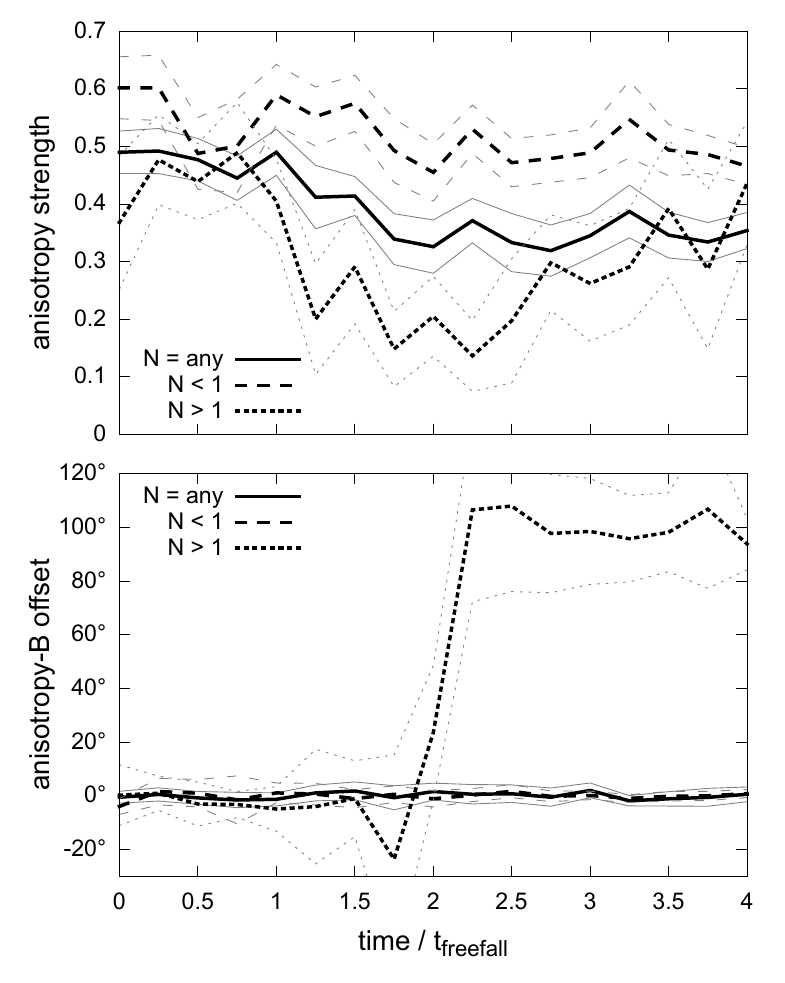}{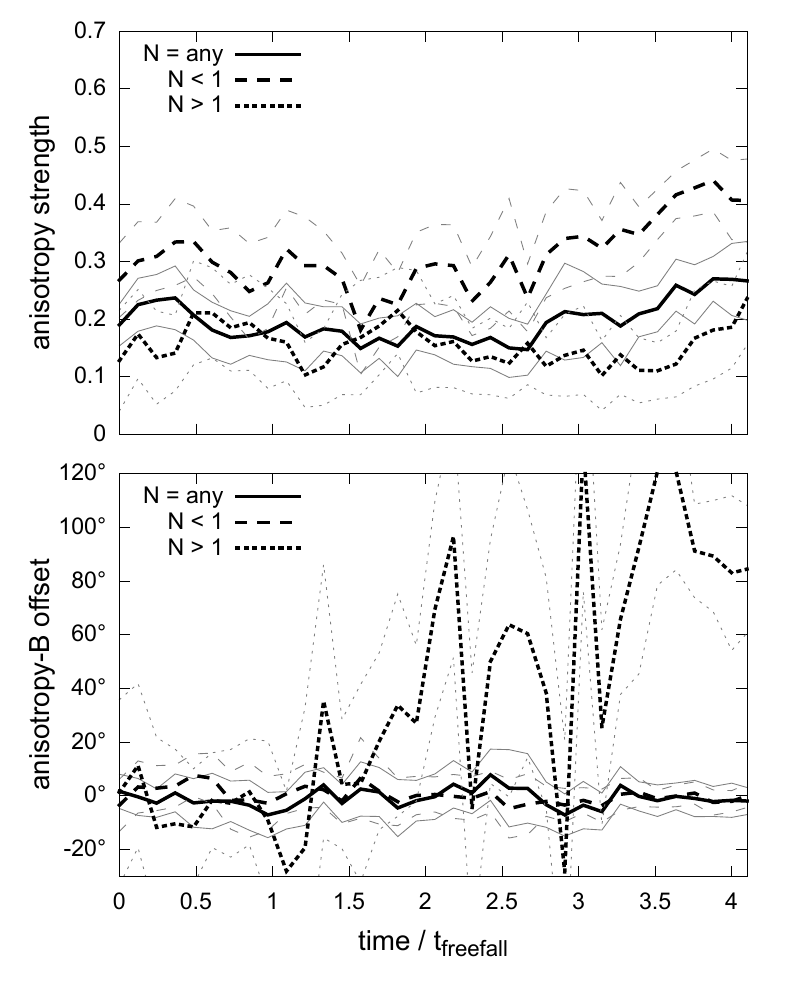}
\caption{%
Development of the anisotropy strength (top panels) and of the offset between
the anisotropy orientation and the large-scale $B$-field (bottom panels) over time.
The anisotropy was measured at the scale $l = 24\px$.
Shown are the results for the B10E1$\zeta$1 simulation with $cr=1$ on the left,
and the B3E1$\zeta$0 simulation with $cr=2$ on the right.
Thick solid lines show the results from analyzing the
complete map, while thick dashed and dotted lines show results from analyzing
only regions with column density $N<1$ and $N>1$, respectively.
The corresponding thin lines indicate the $1\sigma$ uncertainties, obtained
from the disc-sampling procedure (with $D=64\px$; see Section \ref{sec:uncertainty})
and from averaging over the two LOS perpendicular to $\vct{B}$.
\label{fig:tdai}
}
\end{figure*}

Our results for the velocity anisotropy in the face of competition
between turbulence and $B$-fields, but without gravity, are
summarized in Figure \ref{fig:ai-all}.  As we have seen, the
anisotropy is most clearly defined at small scales;
however, small scales are also strongly affected by numerical diffusion.
Hence we have elected to measure it at a scale of
$l = 24\,\text{pixels} \approx 0.08\, L_\text{drv}$.  This scale
is small enough to not be affected by the driving scale, it still exhibits
significant anisotropy for small and medium Mach numbers, and it is big enough to
avoid numerical diffusitiy effects (cf.\ Appendix \ref{app:driving})
as well as possible orientation bias due to the
discreteness of the computational grid.
The figure shows, for each combination of sonic and alfv\'enic Mach number,
the anisotropy strength measured from lines of sight perpendicular and
parallel to the mean $B$-field. For the parallel LOS, all simulations yield a similar
anisotropy strength of $b \approx 0.05$, which coincides with the baseline
anisotropy strength in absence of $B$-fields.  For the perpendicular LOS, 
we systematically observe the anisotropy strength to be larger than this
baseline level, most significantly so for smaller Mach numbers.
Likewise, for perpendicular LOS, we note that the orientation of the
anisotropy coincides very well with the POS $B$-field direction. The
deviation is below $3\dg$ for small Mach numbers and within $20\dg$
for high Mach numbers (cf. Figures \ref{fig:ai-subalf}, \ref{fig:ai-transalf}).

\subsection{Influence of gravity on velocity anisotropy}
\label{sec:res-gravity}

In the second stage of our simulations, we start from a snapshot
with saturated turbulence and turn on gravity while the turbulence
continues to be driven.  The relative strength between the magnetic
forces and gravity is determined by the criticality parameter $cr$
(see Section \ref{sec:simulation}), and we have run the gravitational
stage of our simulations with $cr=1$ (trans-critical cloud)
and $cr=2$ (notably supercritical cloud).
We now investigate whether the oriented structures,
which may form due to gravitational
collapse, have an influence on the velocity anisotropy.

\placefigure{fig:tdai}

Figure \ref{fig:tdai} shows how the velocity anisotropy
(again measured at scale $l=24\px$) develops over time.
$t=0$ is the moment where gravity is switched on, and
we normalize simulation time by the freefall time
$t_\text{ff} = \sqrt{3\pi / 32 G \rho_0}$. 
Here we employ the disc-sampling method
as described in Section \ref{sec:uncertainty}, using discs
of diameter $D=64\px$, to measure the
strength and orientation of the velocity anisotropy and to 
estimate their uncertainties. Additionally we average over the
two lines of sight perpendicular to the mean $B$-field, which
also contributes to the uncertainty estimate.
As was previously indicated in Figures \ref{fig:ai-baseline},
\ref{fig:ai-subalf}, and \ref{fig:ai-transalf}, the disc-sampled
values and error estimates are in good agreement with the
results from averaging over several snapshots.

\begin{figure*}[t]
\epsscale{1.15}
\plottwo{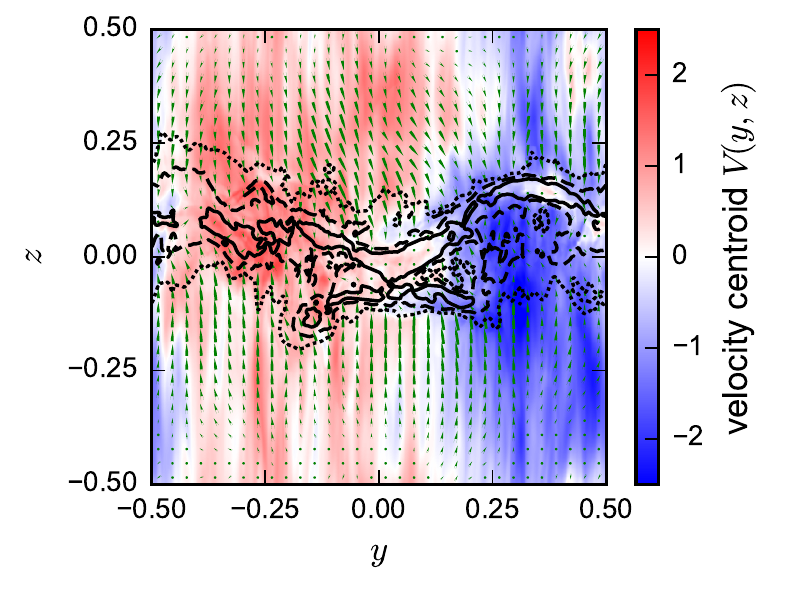}{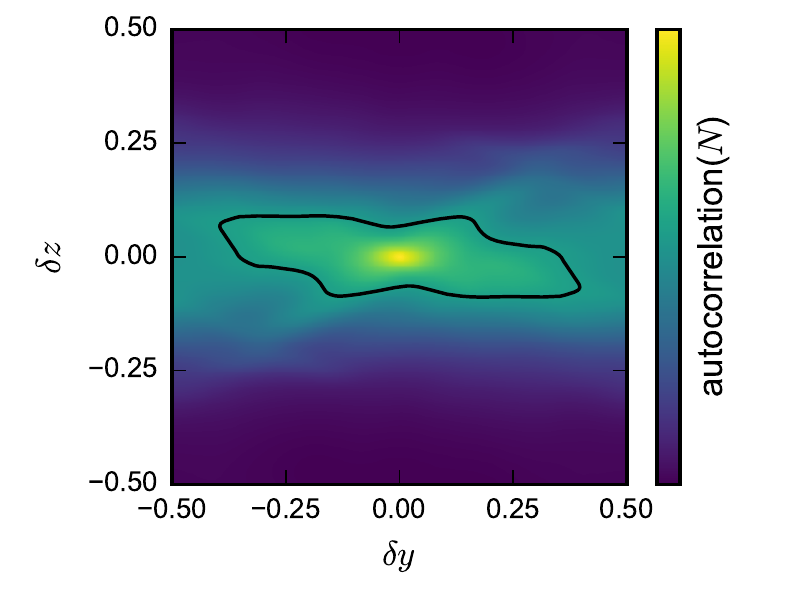}
\caption{%
(Color online)
\emph{Left:}
Map of the velocity centroid (along the $x$-LOS) for the
B10E1$\zeta$1 simulation with criticality 1, at $t = 2\,t_\text{ff}$. 
The contours indicate column densities at 1x (dotted),
2x (dashed), and 4x (solid) the mean column density.
The green arrows show the plane-of-sky velocity field (density-weighted;
length proportional to the magnitude).
\emph{Right:}
Corresponding map of the autocorrelation function of the column density $N$.
The contour surrounds the top 10\% of autocorrelation values.
\label{fig:VCIcr1}
}
\end{figure*}

\begin{figure*}[t]
\epsscale{1.15}
\plottwo{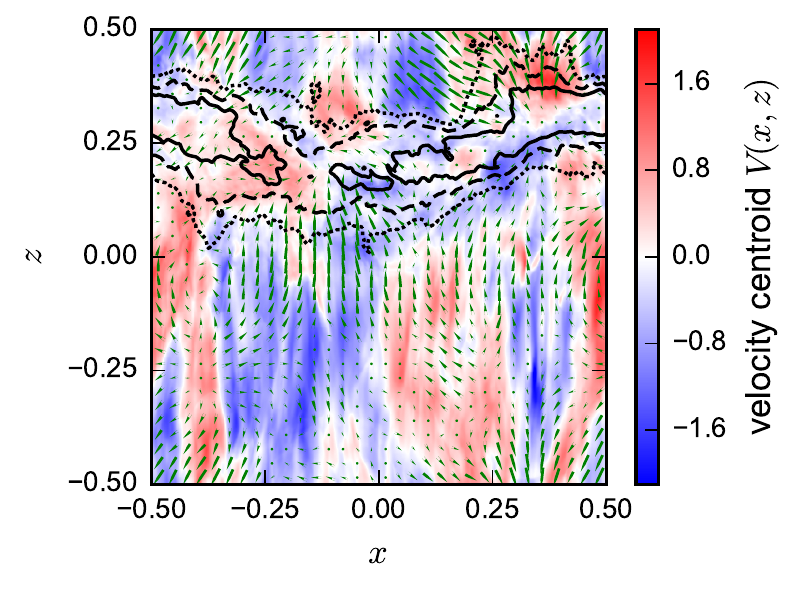}{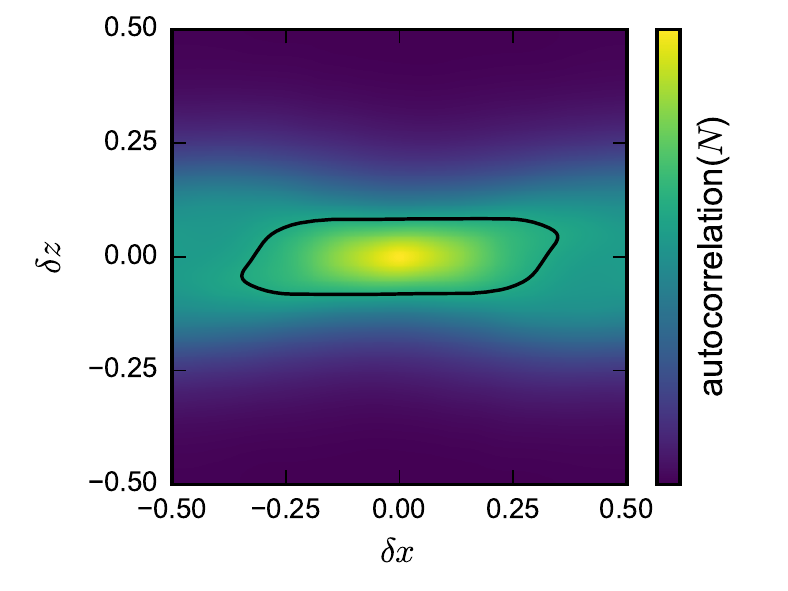}
\caption{%
(Color online)
Like Fig.\,\ref{fig:VCIcr1}, but for the
B3E1$\zeta$0 simulation with criticality 2 at $t = 3.8\,t_\text{ff}$
and along the $y$-LOS.
\label{fig:VCIcr2}
}
\end{figure*}

The lefthand side of Figure \ref{fig:tdai} shows the B10E1$\zeta$1 simulation with $cr=1$ over a period of
four freefall times. When we measure the overall velocity anisotropy for the whole map (these results
are depicted by the solid lines), we note that anisotropy strength drops slightly over time (from
$b \approx 0.5$ to $b \approx 0.35$) but it stays well above baseline levels. Accordingly, the anisotropy
orientation maintains a very tight correlation with the POS $B$-field orientation.
A similar behavior can be observed on the righthand side of Figure \ref{fig:tdai}, which
shows the results for the B3E1$\zeta$0 simulation with $cr=2$, i.e. with weaker $B$-field,
different turbulence driving mode, and higher criticality.  Here the anisotropy maintains
its strength at $b \approx 0.2$, and likewise it aligns within a few degrees with the
POS $B$-field.  From these results alone, one could have the impression that gravity
did not have a big influence on the simulation. 
However, the two simulations represented in Figure \ref{fig:tdai} do exhibit 
obvious formation of extended, filamentary structure due to gravitational collapse,
as can been seen from the maps of the
column (mass) density (indicated by the contours in the lefthand panels of
Figure \ref{fig:VCIcr1} for B10E1$\zeta$1 and of Figure \ref{fig:VCIcr2} for B3E1$\zeta$0).

To investigate whether gravity affects the regions close to and far away from these
filamentary structures, we have divided the map of velocity centroids into two regions:
the \emph{low-density region} where the column density $N$ lies below the mean
column density value (which is $N=1$ in simulation units), and the  \emph{high-density region}
where $N$ is larger than the mean.  We then measured the velocity anisotropy individually
for each region, which is accomplished by restricting the disc-sampling method to discs
which lie exclusively within the selected region.  The results of this investigation are
included in Figure \ref{fig:tdai}.  For the low-density region (which makes up the larger
fraction of the map), the velocity anisotropy behaves similarly to that measured on
the whole map; its strength turns out to be somewhat larger than the whole-map
result, and its alignment with the POS $B$-field is maintained very well throughout
the simulation.  In contrast, the velocity anisotropy measured on the high-density
region undergoes a striking development, which is most prominent for the B10E1$\zeta$1
simulation (left panels in Fig.\ \ref{fig:tdai}): it starts out with a strength close to
that from the low-density region, but then begins to weaken until it reaches almost baseline
levels. At that moment, the orientation of the anisotropy loses its alignment with the
POS $B$-field; it instead starts to align in the \emph{perpendicular} direction, though
with a larger spread ($\pm 20\dg$).
This re-oriented anisotropy then gathers strength and eventually regains the strength
of the initial anisotropy, lying again significantly above the baseline. 
Qualitatively, a similar behaviour is observed for the B3E1$\zeta$0 simulation (right
panels in Fig.\ \ref{fig:tdai}): the velocity anisotropy is initially roughly aligned
with the POS $B$-field, then starts to fluctuate strongly, and eventually settles
at an orientation roughly perpendicular to the $B$-field.

We'd like to stress that the loss and/or change of the velocity anisotropy's alignment
in the high-density regions is indeed caused by gravity, and can not be explained
by e.g.\ shocks which are caused by the turbulence driving.
At the moment where we switch on gravity ($t=0$), the state of the simulation is
identical to the preceeding pure-MHD stage, and Fig.\ \ref{fig:tdai} shows that at
$t=0$ (and in fact for some time thereafter) the velocity anisotropy is well aligned
with the POS $B$-field in all regions, regardless of density.  The loss/change of
alignment is only encountered after the gravitational collapse has begun to form
structures.

\begin{figure*}[tb]
\plottwo{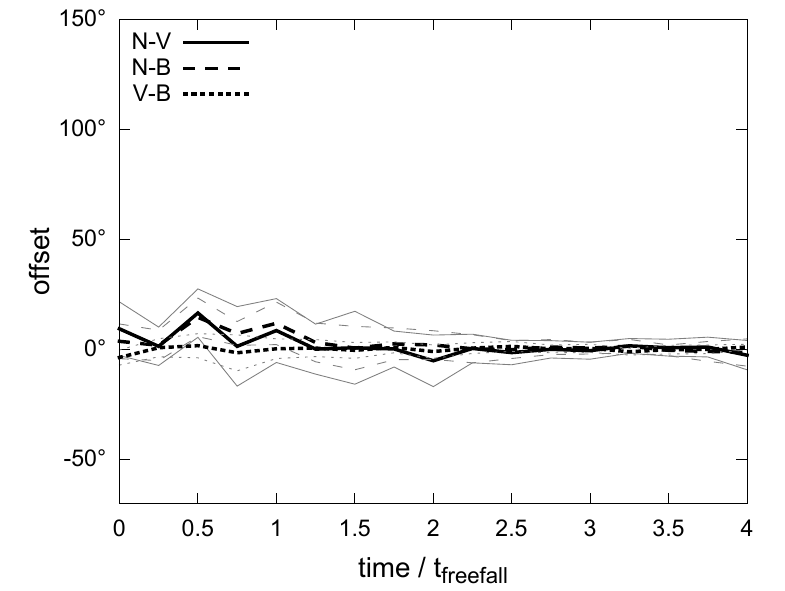}{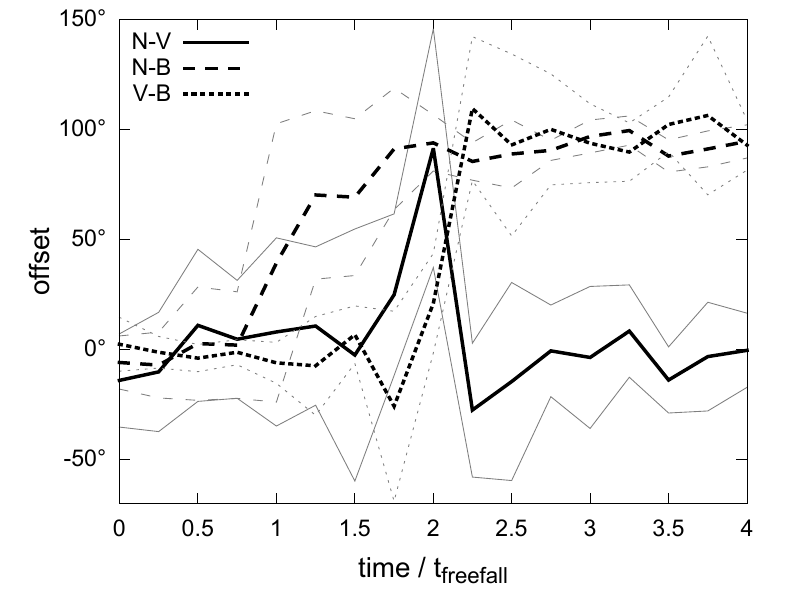}
\caption{%
Development of the offset between the local orientations of the column density ($N$),
velocity anisotropy ($V$), and magnetic field ($B$),
for the B10E1$\zeta$1 simulation with $cr=1$ and starting at the time when
self-gravity is switched on.  The left and right panels show results for the regions with
$N<1$ and $N>1$, respectively.  The thin lines indicate the $1\sigma$ uncertainties as in
Fig.\ \ref{fig:tdai}.
\label{fig:NVB}
}
\end{figure*}

\subsubsection{The correlation between velocity anisotropy and local structure}
\label{sec:vca-vs-local}


\placefigure{fig:NVB}

A possible cause for the changing orientation of the velocity anisotropy in the
high-density regions is that the velocity field is getting affected by
density structures which are emerging due to the gravitational collapse.
To investigate this possibility, we focus on the simulation which exhibits
this reorientation most prominently (B10E1$\zeta$1), and take a look at the
pairwise correlations between the local orientations of the velocity anisotropy ($V$),
the column density ($N$), and the magnetic field ($B$).  Here, the local orientation of the
column density is measured by a method analog to that for the velocity anisotropy
(i.e. based on the the second order structure function of the column density,
cf.\ Sec.\ \ref{sec:analysis}), and analysed at the same scale of $l=24\px$.
The results are shown in Fig.\ \ref{fig:NVB}. On one hand, we observe that in the
low-density region (left panel), all three orientations are parallel to each other
(with offsets $<20\dg$), where the tightest correlation is seen between $V$ and $B$.
Over time, the correlation between all three orientations becomes even stronger.
On the other hand, in the high-density region (see Fig.\ \ref{fig:NVB}, right panel)
the three correlations develop very differently. We observe that first
(around $t=1t_\text{ff}$) the offset between $N$ and $B$ (dashed line) switches from parallel
to perpendicular, and this correlation becomes quite tight ($\sigma < 10\dg$) at
later times. As noted before, the offset between $V$ and $B$ (dotted line) also switches from parallel
to perpendicular, though with some delay compared to the $N-B$ offset. Consequently, the
offset between $N$ and $V$ (solid line) switches briefly from parallel to perpendicular, then back
again.  We note that the correlation between the $V$ orientation and both the $B$ and $N$
orientations is much weaker (with $\sigma \gtrsim 20\dg$) than that between $N$ and $B$.
This indicates that in this simulation, the local structure formation is governed
by the magnetic field.  The fact that the velocity anisotropy only changes its orientation
after the column density structures perpendicular to the $B$-field have formed
(around $t=2t_\text{ff}$) indicates that indeed it is the presence of these structures
which affects the velocity anisotropy orientation.

\begin{figure*}[t]
\epsscale{1.15}
\plotone{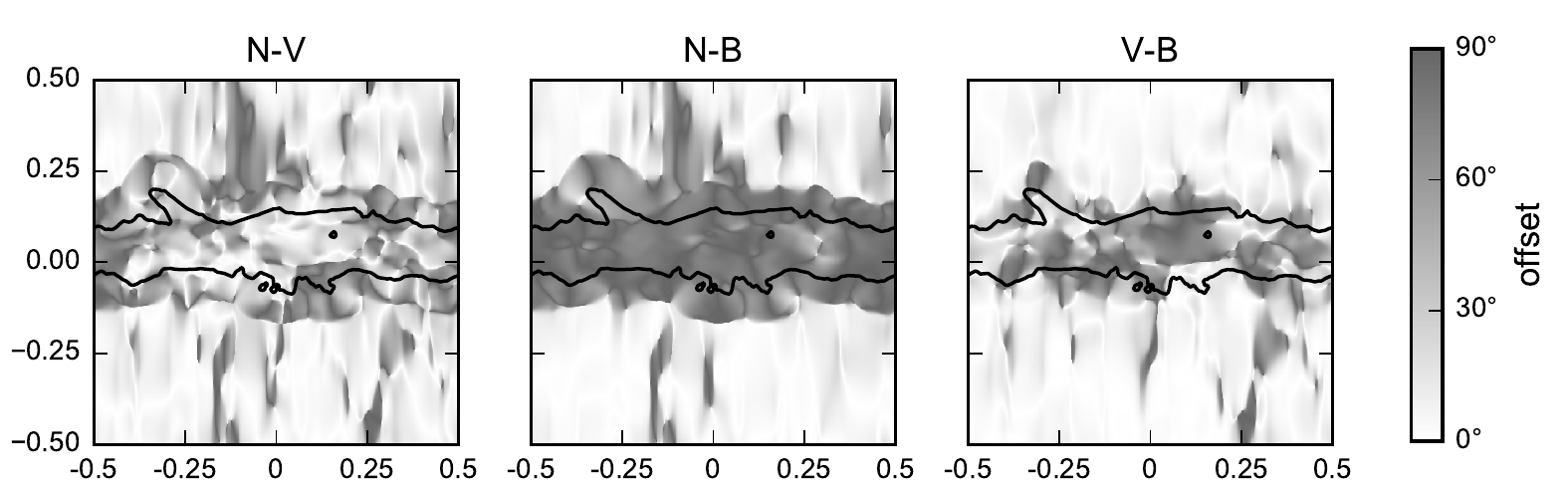}
\caption{%
Maps of the absolute offset between the local orientations of the column density ($N$),
the magnetic field ($B$), and the velocity centroid anisotropy ($V$), for the B10E1$\zeta$1 simulation
with $cr=1$ at $t = 3 t_\text{ff}$, along the $x$-LOS. The black contour indicates $N=1$, i.e.\ the high-density
region is inside this contour while the low-density region is outside.
The mean magnetic field points in the vertical direction ($z$-axis).
\label{fig:offsets-maps}
}
\end{figure*}

\begin{figure}
\epsscale{0.9}
\plotone{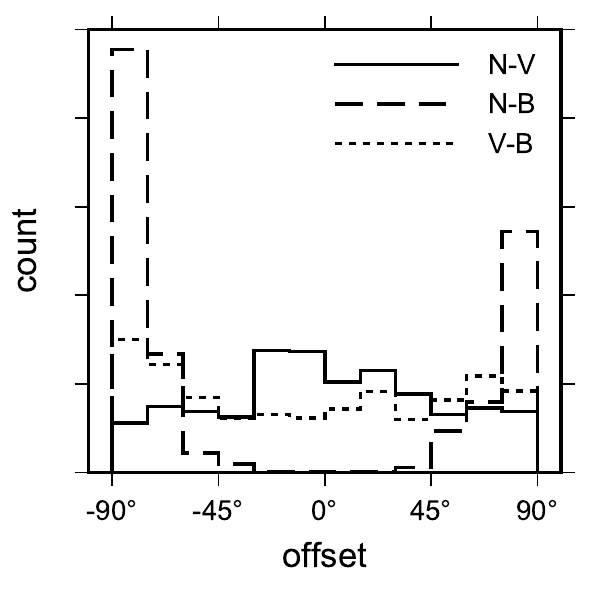}
\caption{%
Histograms of local orientation offsets in the high-density region ($N>1$)
for the same snapshot as in Fig.\,\ref{fig:offsets-maps}.  The histogram
bins are $15\dg$ wide.
\label{fig:offsets-hist}
}
\end{figure}

\placefigure{fig:offsets-maps}
\placefigure{fig:offsets-hist}

Focusing on the later stage of the gravitational collapse (at $t=3t_\text{ff}$
for the aforementioned B10E1$\zeta$1 simulation), Fig.\,\ref{fig:offsets-maps}
shows how the pairwise absolute offsets between the local orientations of 
$N$, $V$, and $B$  are distributed on the plane-of-sky.  We note again that
in the low-density region (i.e.\ outside the indicated $N=1$ contour) all three
orientations tend to be parallel to each other.  The $N-B$ offset (middle panel),
switches very sharply from parallel to perpendicular near the transition from
low- to high-density region.  While the $V-B$ offset (right panel) shows almost exclusive
parallelity in the low-density region, it is dominated by perpendicular orientations
in the high-density region.  The $N-V$ offset presents a more complex pattern, where
perpendicular orientations are observed in the transition region between low and
high column density, but parallel orientations dominate both the diffuse region
outside the cloud and the region deeper inside the cloud.
A more quantitative analysis of the relative orientations between $N$, $V$,
and $B$ inside the high-density region is presented in Fig.\,\ref{fig:offsets-hist}.
The most pronounced feature is seen in the $N-B$ offset, which peaks sharply
at $\pm 90\dg$, indicating a strong tendency for perpendicular orientations.
In contrast, the $V-B$ offset shows a rather flat distribution with weak
peaks at $\pm 90\dg$, i.e.\ a much weaker tendency for perpendicular orientations.
Likewise, the $N-V$ offset displays a shallow peak around $0\dg$, which indicates
a slight preference for parallel orientations.

In summary, we observe that the gravitational collapse has little influence on
the strong correlation between velocity anisotropy and $B$-field orientation in
regions of low column density (they remain parallel), but in higher density regions
only relatively weak correlations can be found for the velocity anisotropy.
There it tentatively aligns parallel to local density structures and hence perpendicular
to the $B$-field, since the structures are strongly influenced by the $B$-field
and form perpendicular to it.  The behavior of the velocity anistropy is particularly
unclear in the transition region between low and high column density (compare Fig.\ \ref{fig:offsets-maps}),
which may obscure its behavior inside the denser region unless the denser region
is of significant extent.  For the simulation discussed in this section (B10E1$\zeta$1, $cr=1$)
this is indeed the case, as the strong $B$-field ($\Mach_A \approx 0.24$) leads to the
formation of an extended filament on the plane-of-sky.  How the velocity anisotropy in
the high density region correlates with such large scale structures in general,
will be discussed in the next subsection.

\subsubsection{The correlation between velocity anisotropy and large-scale structure}
\label{sec:vca-vs-global}

\placefigure{fig:VCIcr1}
\placefigure{fig:VCIcr2}

The two simulations highlighted in Figure \ref{fig:tdai} develop large-scale
column density structures which lie roughly perpendicular to the POS $B$-field (lefthand panels
of Figures \ref{fig:VCIcr1} and \ref{fig:VCIcr2}).  To better quantify how these
structures are oriented, we employ a method based on the autocorrelation
of the column density \citep{li_link_2013}.  Taking into account the periodic boundary
conditions, we first apply a Fourier transform to the entire column density map, then take its
absolute square, then apply an inverse Fourier transform, which yields a map of the
column density's autocorrelation.  We then select the contour which surrounds the
top 10\% of autocorrelation values.  Then we determine the principal axes of the area inside this
contour; the longer axis is used to define the orientation of the structure, and the
aspect ratio (length of longer axis divided by length of shorter axis) can be used to
define how strongly the structure is oriented.  The righthand panels of Figures
\ref{fig:VCIcr1} and \ref{fig:VCIcr2} show the autocorrelation maps with the
top-10\% contour for the two exemplary simulations B10E1$\zeta$1 and
B3E1$\zeta$0, respectively.  For B10E1$\zeta$1, we obtain a POS angle of
$175\dg$ and an aspect ratio of 5.5; for B3E1$\zeta$0, the POS angle is
$3\dg$ and the aspect ratio is 5.1.  That is, in both simulations the column
density structure is strongly oriented, and aligned close to perpendicular to
the mean POS $B$-field (which has a POS angle of $\approx 90\dg$).

As was shown in Fig.\ \ref{fig:tdai},
for the two highlighted simulation runs we find
that the velocity anisotropy in the high-density region
becomes aligned with the orientation of the forming structure, while in the low-density
region the anisotropy stays aligned with the $B$-field.
We remind the reader that the orientation of the
velocity anisotropy is defined as that direction in which the dispersion
of the line-of-sight velocity centroid is minimal. Looking at the map of
the velocity centroid (colormaps in the lefthand panels of Figures
\ref{fig:VCIcr1} and \ref{fig:VCIcr2}), we can get an idea why the anisotropy
orients itself differently in the low- and high-density regions.
In the low-density region, the velociy centroid map exhibits elongated
structures (``striations'') along the $z$-axis (parallel to the $B$-field).
Moving along a striation, the velocity centroid changes little, but moving
across a striation, it changes more rapidly.  Hence the velocity centroid
dispersion is minimal along the $B$-field.
But in the high-density region, the striations disappear, and there is
instead a tendency for the velocity centroid to stay more constant
within individual column density contours. Moving along such a contour
changes the velocity centroid less strongly than moving across the contours.
Hence the velocity centroid dispersion is now minimal along the contour
structures, which lie perpendicular to the $B$-field.

\begin{figure}[tb]
\epsscale{1.2}
\plotone{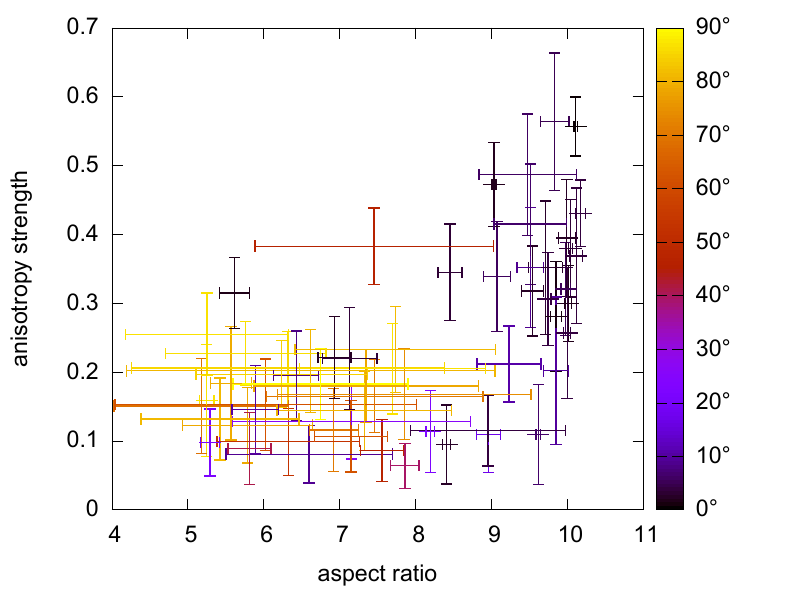}
\caption{%
(Color online)
Correlation between high-density velocity centroid anisotropy and column density autocorrelation structure.
Each data point represents one snapshot from the gravitational stage of one of our
simulations. The horizontal axis shows the aspect ratio of the top-10\% contour of the
column density's autocorrelation (averaged over the two LOS perpendicular to the
$B$-field).  The vertical axis shows the anisotropy strength measured from the
velocity centroid in the high-density region, i.e. where column density $N>1$
(also averaged over the two LOS $\perp \vct{B}$).
The colour indicates the offset between the orientation of the column density
autocorrelation contour and the orientation of the velocity centroid anisotropy.
(This plot only includes snapshots where the aspect ratio is larger than 5, to
prevent the data from becoming too dense.)
\label{fig:all-cdac-vs-vcai}
}
\end{figure}

\placefigure{fig:all-cdac-vs-vcai}

Clearly defined column density structures (where the top-90\%
column density autocorrelation contour has an aspect ratio above 4)
form not only in the two exemplary simulation runs mentioned above, but in about half
(14 of 26) of our simulations with gravity (see Table \ref{tab:list-of-sims} for details).  
Figure \ref{fig:all-cdac-vs-vcai} shows how these structures correlate with
the averaged velocity anisotropy in the high-density region. 
The yellow-colored data points correspond to snapshots from early stages
of gravitational collapse, where structures have started to form (lower aspect ratio)
perpendicular to the $B$-field, and the velocity anisotropy has weakened but
still tends to be aligned with the $B$-field, hence having $\approx 90\dg$ offset
from the orientation of the density structure.
In the progression from orange over red to purple data points, the structures
become more pronounced (increasing aspect ratio) while the velocity anisotropy
weakens and starts to deviate from the $B$-field orientation.
The black data points correspond to snapshots from the final stage of gravitational collapse,
where structures are strongly oriented (aspect ratio $>8$) and the velocity anisotropy has
become aligned parallel to these structures (within $10\dg$) and has gained strength.

In summary, for those simulations where gravitational collapse leads to clearly
oriented structures (like filaments), we universally find that the velocity anisotropy
in the high-density region becomes preferentially aligned with the structure instead of with the
$B$-field.  However, as pointed out in subsection \ref{sec:vca-vs-local},
this correlation is weaker than what is found in the low-density region,
where velocity anisotropy strongly aligns parallel to the $B$-field.

\section{Discussion}
\label{sec:discussion}

As detailed in Section \ref{sec:intro}, velocity anisotropy has been
studied before by a variety of means, and here we wish to discuss
similarities and differences with the results of some of the previous works.
Most closely related to the present investigation are the studies
from \citet{esquivel_velocity_2011} (hereafter EL11) and \citet{burkhart_measuring_2014}
which also employ velocity centroid data (obtained from MHD simulations
of turbulence without self-gravity) to measure the velocity anisotropy.
In these studies, the anisotropy strength is measured using the
``isotropy degree'' which is the ratio of the velocity centroid's structure
function (at a certain scale) in the directions
parallel and perpendicular to the POS $B$-field.
These studies did not discuss how the orientation of the anisotropy
could be determined, and assumed that it aligns with the POS $B$-field
(at least on small scales).
In contrast, our method of fitting the angular dependence of the
structure function (at a certain scale) to a simple model function 
directly measures not only the strength of the anisotropy, but also
its orientation.  Hence our method is of advantage if one wants to apply
it to observational data, where the POS $B$-field orientation is
not necessarily known (or not necessarily with good accuracy).

Regardless of the different anisotropy measure, we confirm one
of the major findings of EL11, namely that
the anisotropy strength depends both on the alfv\'enic and the sonic
Mach number (recall Figure \ref{fig:ai-all} for our results). In general,
the anisotropy is most pronounced if both Mach numbers are small.
Increasing the alfv\'enic or the sonic Mach number leads to a decrease
of the anisotropy strength, though we find that the anisotropy remains
detectable even for highly supersonic ($\Mach \approx 17$) or for mildly
super-alfv\'enic ($\Mach_A \approx 1.2$) conditions. Likewise,
\citet{esquivel_velocity_2011} find detectable anisotropy up
to $\Mach_A \approx 1.5$.

\begin{figure*}[t]
\epsscale{1.2}
\plotone{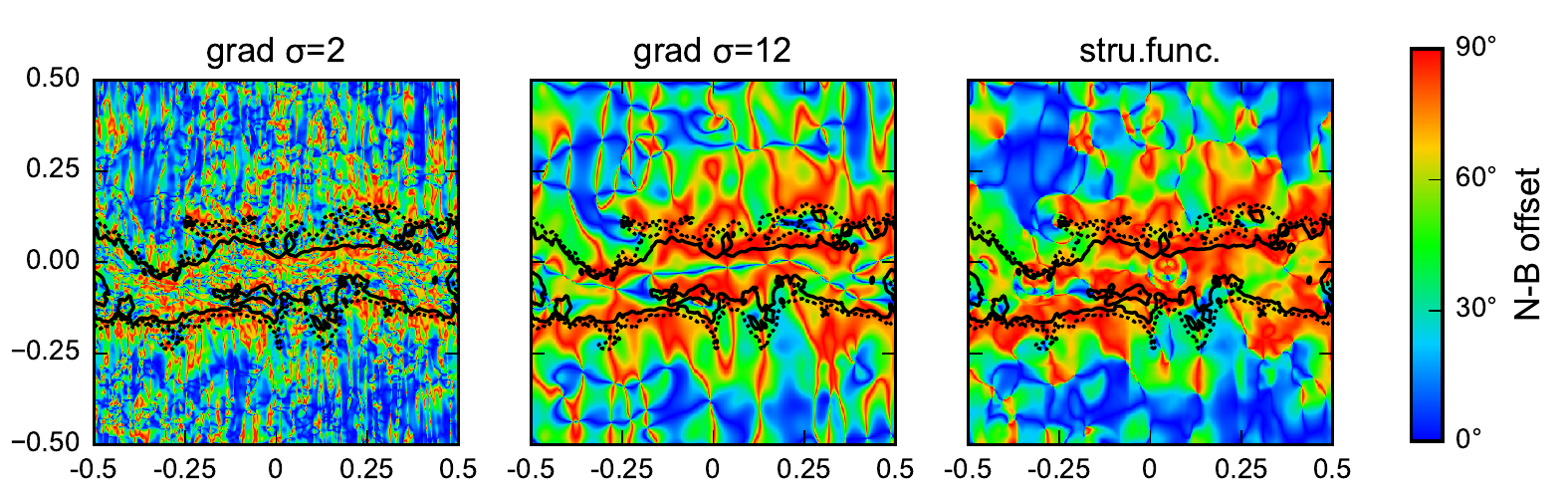}
\caption{%
(Color online)
Map of the absolute offset between the local orientations of column density ($N$)
and magnetic field ($B$), for the B10E2$\zeta$0 simulation with $cr=1$ at $t = 4\,t_\text{ff}$,
along the $x$-LOS, where the $B$-field points in the vertical direction ($z$-axis).
The dotted and solid black contours indicates $N=1$ and $N=2$, respectively
(in units of mean column density).
The three panels show different methods for determining the local $N$-orientation.
In the left panel, the density gradient method is employed, with width $\sigma=2\px$ for
the Gaussian derivative kernel (see text); likewise in the middle panel with $\sigma=12\px$.
In the right panel, local $N$-orientation is determined from the anisotropy of the 
local structure function of $N$, at the scale $l=24\px$.
The colour scale has been chosen to allow visual comparison with Fig.\ 9
from \cite{planck_collaboration_planck_2016}.
\label{fig:B10E2}
}
\end{figure*}

Measuring the velocity anisotropy from observational data has been
the topic of \cite{heyer_magnetically_2008} (hereafter H+08) and \cite{heyer_trans-alfvenic_2012}
(hereafter HB12), with the goal of estimating the magnetic field strength
in the Taurus molecular cloud.  These studies used a more elaborate method
than the one presented here, which applies principal component analyis to
PPV spectroscopic data.  By calibrating the method against numerical simulations, the strength
of the POS $B$-field component can be inferred.
H+08 focused on low-density striations in the envelope of the cloud,
and report a well-defined velocity anisotropy which is aligned with the POS $B$-field.
Their PCA-based method yields an estimate of $B_\text{POS} = 14\,\muG$,
which coincides well with the estimate from the Chandrasekhar-Fermi method.
HB12 extended the study of the Taurus cloud to higher density regions
through inclusion of optically thinner $^{13}\text{CO}$ data.  Notably, no
significant velocity anisotropy was detected in the higher density regions
(where visual extinction $A_V > 2$).  HB12 propose that the absence of
velocity anisotropy in these regions is due to the transition from sub- to
super-alfv\'enic conditions with increasing density, although their major
argument for the existence of super-alfv\'enic turbulence in the high-density
regions seems to be the absence of velocity anisotropy.  HB12 further suggest
that ambipolar diffusion (i.e. the contracting gas decouples from the $B$-field
and leaves the $B$-field behind) may be responsible for this change of the
alfv\'enic Mach number.

Our study offers an alternative explanation for the observations in HB12.
As we saw in Section \ref{sec:res-gravity}, if the gravitational contraction
causes the formation of oriented structures (which certainly applies to
the filamentary Taurus cloud), then the velocity anisotropy in the high-density
region may disappear, or even become aligned with the structure instead of
with the $B$-field.  Hence the absence of velocity anisotropy in the
$A_V > 2$ $^{13}\text{CO}$ data may simply be caused by gravity,
and is not necessarily an indication for super-alfv\'enic conditions inside
the high-density regions.


\placefigure{fig:B10E2}

In fact, we can scale one of our simulations (which were carried out in
reduced units, see Section \ref{sec:simulation}) such that the physical
parameters match the conditions in the Taurus cloud (as reported in
H+08 and HB12) reasonably well.  For a discussion about how the
dimensional scaling is achieved, see e.g. \citet[Section 4.1]{li_magnetized_2015}.
Our simulation B10E2$\zeta$0 with $cr=1$ possesses an alfv\'enic Mach number
of $\Mach_A \approx 0.37$ before gravity is switched on (cf. Table \ref{tab:list-of-sims}),
which is close to the $\Mach_A \approx 0.5$ reported in H+08.  Using an
isothermal temperature of $T=15\,\text{K}$ (from H+08) and choosing the
box size as $L = 9.2\,\text{pc}$ (somewhat larger than the $3.5\,\text{pc}$
region investigated in HB12), the dimensional scaling yields a mean
$B$-field strength of $14\,\muG$, matching the estimate from H+08.
The total mass inside this scaled simulation is $\sim 3400 M_\odot$,
and the mean column density is $N(\text{H}_2) = 1.94 \cdot 10^{21}\,\text{cm}^{-2}$
which corresponds to a visual extinction of $A_V = 1.04\,\text{mag}$
\citep{bohlin_survey_1978, vrba_ratio_1984}.
After $t = 4 t_\text{ff} = 15.5\,\text{Myr}$ of gravitational collapse, a
well-defined filamentary structure has formed in this simulation, which
is nearly perpendicular to the mean $B$-field (offset $91\dg \pm 2\dg$).
See the black contours in Fig.\ \ref{fig:B10E2} for an illustration of this structure.
In the low-density region (here using $A_V < 2$ to match HB12), we measure a
velocity anisotropy strength $b = 0.59 \pm 0.04$ which is tightly aligned with
the $B$-field (deviation $< 2\dg$).  In the high-density region ($A_V > 2$), the
anisotropy is much weaker at $b = 0.2 \pm 0.1$ and possesses a very unclear
alignment (POS angle $147\dg \pm 48\dg$).  This matches HB12's observation
of an inconclusively defined anisotropy in the high-density region.
However, as our simulation provides full access to the three-dimensional
velocity field, we can evaluate the Mach numbers inside the low- and
high-density region.  First we note that the $B$-field strength doesn't
deviate strongly from the mean value due to the sub-alfv\'enic setup of this
simulation.  Thus we obtain $\Mach_A = 0.56$ in the region with $A_V<2$
and $\Mach_A = 0.54$ for $A_V>2$, for both lines of sight perpendicular
to the mean $B$-field.  While the Mach number has increased
compared to the initial condition (which is due to energy released by the
gravitational collapse), we don't find an appreciable difference in Mach
numbers between the low- and high-density region.
Nevertheless, the explanation from HB12 which invokes ambipolar
diffusion cannot be ruled out by our simulation, as our numerical
method employs ideal MHD and hence does not include ambipolar diffusion effects.

Although the focus of this study has been on the orientation of the velocity anisotropy,
we would like to add some final remarks on the relation between the
orientations of the $B$-field and of the density structures. In the simulations
presented here, we found large-scale dense structures forming perpendicular to the mean
(i.e.\ large-scale) $B$-field.
In observations of nearby molecular clouds from the Gould Belt, it has been found that the large-scale
cloud structure aligns preferentially either perpendicular or parallel to the
large-scale (inter-cloud) plane-of-sky $B$-field, giving rise to a bimodal
distribution of the cloud-field alignment \citep{li_link_2013}.
We'd like to stress that, in the present simulations, we do not encounter this bimodal alignment for
the large-scale structures, which may be due to the limited scope and restricted physics model of
our simulation setup.
However, we do see a bimodal alignment between density structures and $B$-field
on the \emph{local} scale -- they tend to be parallel in regions of low column density,
but perpendicular for high column density.
This matches observational findings from Gould Belt clouds
reported in \cite{planck_collaboration_planck_2016} (herafter: Planck XXXV).
While we determine the orientation of local structures from the anisotropy of
the column density's local structure function at a given scale $l$,
in Planck XXXV the local structure orientation is defined
as perpendicular to the gradient of the column density.
Computing this gradient using a Gaussian derivative kernel \citep{soler_imprint_2013}
also allows to set the scale at which the local orientation is analyzed
(this corresponds to smoothing the column density map with a Gaussian of
a certain width $\sigma$ before taking the gradient).
To test whether the differing definitions of local structure orientation 
influence the conclusion about the local structure-field alignment,
we return to the previously discussed B10E2$\zeta$0 simulation, since
(as argued before) it has comparable properties to the Taurus region,
which was also part of the Planck XXXV study.
Fig.\ \ref{fig:B10E2} presents the plane-of-sky distributions of the
local structure-field alignment, using the density gradient method with
convolution widths $\sigma=2\px$ and $\sigma=12\px$ (left and middle
panel, respectively) and our structure function based method with
$l=24\px$ (right panel).
First, we note that the alignment patterns in the left and middle panel
possess strong visual similarity to those in the $10^\prime$ and $60^\prime$
maps of the Taurus region in Fig.\ 9 from Planck XXXV.
Second, we note that in the middle and right panel, the size and distribution of
the patches with parallel (blue) and perpendicular (red) alignment
closely match, indicating that these two analysis agree both in the
analysed scale and in the result for the local structure orientation.\footnote{%
The density gradient method tends to exhibit artefacts where the
column density displays a ridge or a valley, which accounts for the
major part of the discrepancies between the middle and right panel
of Fig.\ \ref{fig:B10E2}.}  Regardless of the analysis method or scale,
we observe that parallel structure-field alignment occurs mostly in the
low-density region, while the high-density region is dominated by perpendicular
alignment, i.e.\ the same bimodal distribution of alignment as reported in
Planck XXXV.  But to reiterate, this bimodal alignment is here seen on the
sub-cloud scale and must not be confused with the bimodal alignment seen
on the cloud/inter-cloud scale in \cite{li_link_2013}.

\section{Summary}

Velocity anisotropy is present in synthetic maps of line-of-sight velocity centroid
data which we have obtained from numerical simulations of driven
magnetohydrodynamics turbulence.  The two-point second-order structure
function of the velocity centroid proves to be a useful tool for detecting
and quantifying this anisotropy.  In the sub- to trans-alfv\'enic regime which
we have investigated here, we reliably find significant velocity anisotropies
which are strongly aligned with the plane-of-sky (POS) magnetic field.  This holds
particularly when the anisotropy is evaluated at scales which are small
compared to the turbulence driving scale.  Additionally we find that the
turbulence driving mechanism (here modelled as either purely solenoidal
or purely compressive) has a minor to insignifcant influence on the
quantitative results for the velocity anisotropy.
Hence the method employed here may be used to infer the orientation
of the POS magnetic field from velocity data, coming e.g. from spectroscopic observations.

When self-gravity is neglected, we find that the strength of the velocity
anisotropy depends on both the sonic and the alfv\'enic Mach number
($\Mach$ and $\Mach_A$, respectively), in good agreement with
\citet{esquivel_velocity_2011}. Hence a determination of the anisotropy
strength is not sufficient to determine $\Mach_A$, even if the $B$-field
is known to lie on the plane-of-sky.  However, when other
observations or constraints for $\Mach$ (e.g. from non-thermal linewidths)
and for $B_\text{LOS}$ (e.g. from Zeeman observations) are available, it is possible to
infer additional constraints for $\Mach_A$ through the velocity anisotropy.

When self-gravity is taken into account, we find that significant velocity
anisotropy is still present in regions of lower column density, where it
remains aligned with the POS $B$-field.  In contrast, in regions of higher
column density, the velocity anisotropy may disappear, or there might
even develop an anisotropy which is aligned not parallel to the POS
$B$-field, but parallel to the density structures which emerge from the
gravitational collapse.  This holds particularly if these structures are of
extended size and exhibit clear directional preference, like e.g. filaments.

This result provides an explanation for the loss of velocity anisotropy 
which has been observed in high-density regions of the Taurus molecular
cloud \citep{heyer_trans-alfvenic_2012}.  It has been proposed that this
loss is caused by the transition to super-alfv\'enic conditions inside the
high-density region, but in our simulation (with parameters matching
the Taurus cloud) we find that the alfv\'enic Mach number in the low-
and high-density regions is the same, $\Mach_A \approx 0.55$.
Nevertheless this simulation also
exhibits the observed loss of velocity anisotropy in the high-density
region.  Hence we conclude that this loss is not necessarily associated
with a transition to super-alfv\'enic  conditions.

We expect that our method for investigating the velocity anisotropy
can be applied to observational velocity data in a fairly straightforward
manner. While projection effects certainly need to be taken into
account, \citet{burkhart_measuring_2014} have shown that they do
not preclude the anisotropy from being detected in velocity
centroid data.  Another possible caveat is determining a good
column density threshold for separating
the low- from the high-density region, as the velocity anisotropy may
behave differently in these regions. The simple criterium of using the
mean column density, as used for our simulation data, can not
directly be applied to observations, since the mean column density
changes when the studied region is shrunk or enlarged . We plan to investigate this
issue and apply our method to observational data in a future work.

\acknowledgements

This research was supported by
the Hong Kong Research Grants Council, project ECS 24300314;
and by the CUHK Direct Grant for Research Project 4053073 and 4053126.
WJ gratefully acknowledges financial support from the Hong Kong Research Grants Council
Collaborative Research Fund ``Research in Fundamental Physics: from the Large
Hadron Collider to the Universe''.

\appendix

\section{Turbulence Driving Scheme}
\label{app:driving}

In our simulations, turbulence is driven by periodically exciting random
velocity perturbations in Fourier space. This follows the scheme set forth
in \citep{stone_dissipation_1998}. When a certain period $dt_{\text{drv}}$
of simulation time has passed, we set up a field of velocity perturbations
$\vct{a}(\vct{k})$, where $\vct{k}$ is the wave vector in Fourier space,
such that each component $a_i(\vct{k})$ is normally distributed with mean
zero and variance $\propto k^6 \exp(-8 k / k_{\text{drv})}$.
This ensures that the orientations of the vectors $\vct{a}(\vct{k})$ are
isotropically distributed, and that the perturbation power spectrum
(accounting for the degeneracy in $k$) follows
\begin{equation}
k^2 | \vct{a}(\vct{k}) |^2 \,\propto\, k^8 \exp(-8 k / k_{\text{drv}}) \quad.
\end{equation}
This power spectrum is sharply peaked around $k = k_{\text{drv}}$,
hence the spatial driving scale is given
by $L/k_{\text{drv}}$, where $L$ is the box length of the simulation domain.

Following \citep{schmidt_numerical_2006},
to allow for a controlled mix between solenoidal and compressive components
in the perturbing field, we apply a $\vct{k}$-dependent projection matrix
$\mtx{P}(\vct{k})$ to the generated field $\vct{a}(\vct{k})$:
\begin{align}
\delta\tilde{\vct{v}}(\vct{k}) &= \mtx{P}(\vct{k}) \vct{a}(\vct{k})
\\
P_{ij}(\vct{k}) &= \zeta \delta_{ij} \,+\, (1 - 2 \zeta) k_i k_j / |\vct{k}|^2 
\end{align}
The parameter $\zeta \in [0,1]$ determines the fraction of solenoidal components.
For $\zeta=1$, the driving field is purely solenoidal (divergence-free, non-compressive),
which can be pictured as turbulence driven by stirring motions. On the other hand,
for $\zeta=0$, the driving field is purely compressive (curl-free), which can be
pictured as turbulence driven by compressing or dilating motions (e.g. shock waves).

\begin{figure}[t]
\plotone{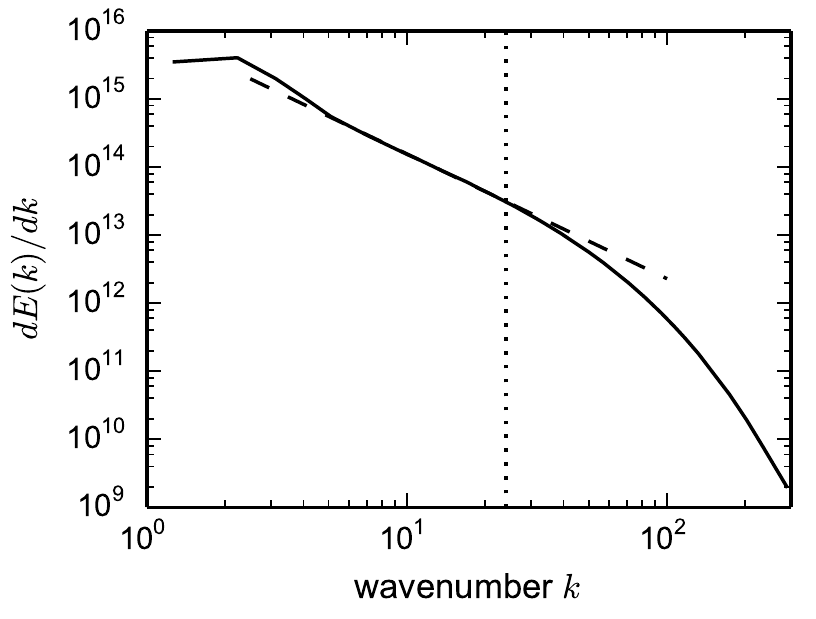}
\caption{%
Turbulent power spectrum of the B0E1$\zeta$1 simulation.
The dashed line shows a fit to the spectrum inside the inertial range
($5 \lesssim k \lesssim 25$), extended for better visibility.
The dotted line at $k=24$ indicates the spatial scale $l=N/k=24\px$,
where $N=576\px$ is the number of grid points per dimension.
\label{fig:turbspect}
}
\end{figure}

Each component $\delta \tilde{v}_i(\vct{k})$ is then multiplied with a uniformly random complex phase
(under the provision that $\delta \vct{v}(-\vct{k}) = \delta \vct{v}(\vct{k})^*$ to ensure
that its Fourier transform is real-valued),
which avoids spatially static patterns in the driving field. The Fourier-transformed velocity
perturbations $\delta \vct{v}(\vct{x})$ are then shifted such that no net momentum will be added
to the simulation domain, and normalized such that the increase in total kinetic energy will match
a prescribed parameter value, $d E_{\text{drv}}$.  In effect, the turbulent energy input ratio,
$d E_{\text{drv}} / dt_{\text{drv}}$, will be constant over time.

The turbulent power spectrum is given by $dE(k)/dk$, where $E(k)$ is the
total specific kinetic energy at scales with wavenumbers up to $k$, i.e.
$E(k) = \int_{|\vct{k}'| \leq k} d^3\vct{k}' \tilde{\vct{v}}(\vct{k}')^2$.
For the B0E1$\zeta$1 simulation, this power spectrum is shown in
Fig.\,\ref{fig:turbspect}; here the spectrum was averaged over a
series of ten snapshots of fully developed turbulence. As our employed
MHD code is based on finite differencing, it is rather diffusive, so that
the turbulent inertial range is not well developed; it spans the range
$5 \lesssim k \lesssim 25$.  In Section \ref{sec:results} some analyses are
carried out at ``small'' spatial scales; we chose to employ the scale
$l=24\px$ for this which, as Fig.\,\ref{fig:turbspect} shows, still lies
inside the inertial range and is hence unaffected by numerical dissipation effects.


\bibliographystyle{aasjournal}
\bibliography{refs}

\end{document}